\documentclass[aps,prd,twocolumn,10pt,nofootinbib,superscriptaddress,floatfix]{revtex4-2}

\usepackage{amsmath,amssymb,bm,mathtools}
\usepackage{graphicx,booktabs}
\usepackage[caption=false]{subfig}
\usepackage[normalem]{ulem}
\DeclareUnicodeCharacter{2212}{-}
\usepackage{xcolor}

\definecolor{apsblue}{rgb}{0.18,0.19,0.57} 
\usepackage[
    colorlinks=true,
    linkcolor=apsblue,
    citecolor=apsblue,
    urlcolor=apsblue,
    pdfborder={0 0 0}
]{hyperref}
\usepackage{siunitx}
\usepackage{float}
\usepackage{orcidlink}

\begin{document}

\title{Suppression of $^{14}\mathrm{C}$ photon hits in large liquid scintillator detectors via spatiotemporal deep learning}

\author{Junle Li\orcidlink{0009-0003-2702-4483}}
\affiliation{School of Aeronautics and Astronautics, Zhejiang University, Hangzhou 310027, China}

\author{Zhaoxiang Wu\orcidlink{0009-0001-1308-9505}}
\affiliation{Institute of High Energy Physics, Chinese Academy of Sciences, Beijing 100049, China}
\affiliation{School of Physical Sciences, University of Chinese Academy of Sciences, Beijing 100049, China}

\author{Guanda Gong\orcidlink{0000-0001-7192-1833}}
\affiliation{Institute of High Energy Physics, Chinese Academy of Sciences, Beijing 100049, China}

\author{Zhaohan Li\orcidlink{0009-0007-8622-728X}}
\email{Corresponding author: lizhaohan@ihep.ac.cn}
\affiliation{Institute of High Energy Physics, Chinese Academy of Sciences, Beijing 100049, China}

\author{Wuming Luo\orcidlink{0000-0003-0137-1797}}
\email{Corresponding author: luowm@ihep.ac.cn}
\affiliation{Institute of High Energy Physics, Chinese Academy of Sciences, Beijing 100049, China}

\author{Jiahui Wei\orcidlink{0000-0001-5230-467X}}
\affiliation{Institute of High Energy Physics, Chinese Academy of Sciences, Beijing 100049, China}

\author{Wenxing Fang\orcidlink{0000-0002-5247-3833}}
\affiliation{Institute of High Energy Physics, Chinese Academy of Sciences, Beijing 100049, China}

\author{Hehe Fan\orcidlink{0000-0001-9572-2345}}
\email{Corresponding author: hehefan@zju.edu.cn}
\affiliation{College of Artificial Intelligence, Zhejiang University, Hangzhou 310027, China}

\date{October 16, 2026}

\begin{abstract}
Liquid scintillator detectors are widely used in neutrino experiments due to their low energy threshold and high energy resolution. 
Despite the tiny abundance of $^{14}$C in LS, the photons induced by the $\beta$ decay of the $^{14}$C isotope inevitably contaminate the signal, degrading the energy resolution. In this work, we propose three models to tag $^{14}$C photon hits in e$^+$  events with $^{14}$C pile-up, thereby suppressing its impact on the energy resolution at the hit level: a gated spatiotemporal graph neural network and two Transformer-based models with scalar and vector charge encoding. 
For a simulated test dataset in which each event contains one $^{14}\mathrm{C}$ and one e$^+$ with kinetic energy below 5~MeV, the models achieve $^{14}\mathrm{C}$ recall rates of 25\%–48\% while maintaining $\mathrm{e}^{+}$ to $^{14}\mathrm{C}$ misidentification below 1\%, leading to a large improvement in the resolution of total charge for events where $\mathrm{e}^{+}$ and $^{14}\mathrm{C}$ photon hits strongly overlap in space and time. 
\end{abstract}

\maketitle

% \linenumbers

\section{Introduction}

\label{sec:introduction}

Pile-up, namely the overlap of signal and background processes within the same data acquisition window, is a well-known challenge in particle physics experiments. 
It can induce fake signal yields or degrade detector resolution~\cite{BOREXINO:2014pcl,ATLAS:2014riz,CMS:2014xym,SNO:2015wyx, KamLand:Obara:2022jhg}, 
thereby affecting physics sensitivity. 
With the advantages of low energy threshold and high energy resolution, liquid scintillator (LS) detectors~\cite{Borexino:2007kvk,KamLAND:2002uet,Lozza:2012bf, DayaBay:2022orm,RENO:2012mkc,DoubleChooz:2011ymz} are widely used in neutrino experiments. 
Regardless of the LS recipe, carbon atoms are usually the dominant component~\cite{JUNOLS:JUNO:2020bcl}.
Despite its tiny abundance in LS, the $^{14}$C isotope decays and induces scintillation photons. When these photons fall within the same data acquisition window as a signal particle, they inevitably contaminate the signal photons. This so-called $^{14}$C pile-up effect worsens the energy resolution or introduces background and is a common challenge for all LS detectors~\cite{BOREXINO:2014pcl,SNO:2015wyx,JUNO:2021vlw,Chen:2023xhj}. 
This effect becomes even more critical for large-scale detectors such as the Jiangmen Underground Neutrino Observatory (JUNO)~\cite{JUNO:2021vlw, Fang:2026cvt}, where the large target mass results in a more frequent occurrence of pile-up events. 
Meanwhile the energy resolution is crucial for high-precision neutrino oscillation measurements.
While hit-level pile-up suppression has been extensively developed for jet reconstruction at the Large Hadron Collider (LHC)~\cite{Vaughan:2025ijq,Algren:2024bqw,PuppiML:Li:2022omf,ArjonaMartinez:2018eah,Puppi:Bertolini:2014bba,Soyez:2012}, 
comparable approaches are less explored in LS detectors. 
In this paper, we investigate the feasibility of utilizing deep learning methods to identify and suppress $^{14}\mathrm{C}$ photon hits in order to mitigate their impact on energy resolution. We focus specifically on $^{14}\mathrm{C}$ pile-up events that contain only a single $^{14}\mathrm{C}$ pile-up, given that multiple $^{14}\mathrm{C}$ pile-up has a much lower probability of occurring and a more complex event topology. Meanwhile the pile-up contamination from radioactive backgrounds or cosmogenic isotopes can be safely ignored, since their rates are at least three orders of magnitude lower than $^{14}\mathrm{C}$.
The JUNO experiment is used as a representative case study.

JUNO is a 20-kton LS detector designed primarily to determine the neutrino mass ordering, and it requires an unprecedented energy resolution of $3\%$ at $1\,\mathrm{MeV}$~\cite{JUNO:2015zny}. 
This performance relies on high optical coverage, provided by approximately 17,600 20-inch and 25,600 3-inch photomultiplier tubes (PMTs)~\cite{JUNO:2024fdc,JUNO:2021vlw}, an optimized LS recipe to achieve high light yield as well as excellent transparency~\cite{JUNO:2025fpc}, extensive PMT R\&D to increase the quantum efficiency for photon detection, and advanced reconstruction algorithms~\cite{JUNO:2024fdc,Huang:2022zum,Takenaka:2025hgi,Qian:2021vnh}. Detailed information on JUNO can be found in Refs.~\cite{JUNO:2025fpc,JUNO:2021vlw}.

JUNO detects reactor $\bar{\nu}_e$ via inverse beta decay (IBD), $\bar{\nu}_e + p \rightarrow n + e^{+}$, 
the average deposited energy of a positron is at the MeV scale. 
In contrast, the peak  and endpoint energies of the $^{14}$C beta decay are  $0.04\,\mathrm{MeV}$ and $0.16\,\mathrm{MeV}$, respectively, yielding substantially fewer scintillation photons. 
As a result, when the two processes overlap in time, the $^{14}$C photon contribution is embedded in the dominant positron signal and becomes difficult to distinguish at the hit level.

\begin{figure*}[t]
    \centering
    \subfloat[Large separation ($\Delta t = 455.0~\mathrm{ns}$)\label{fig:class_count_dt_large}]{
        \includegraphics[width=0.48\textwidth]{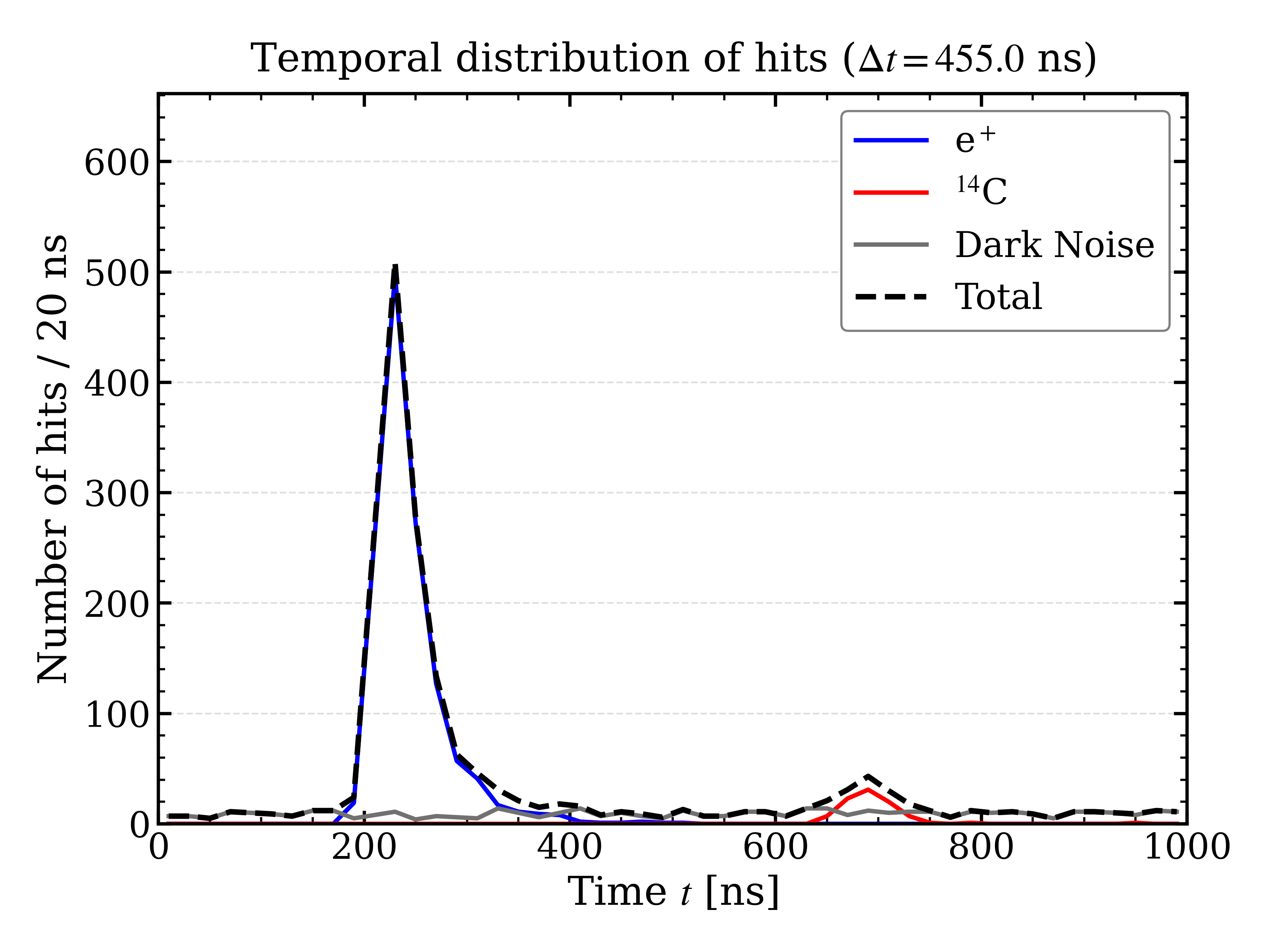}
    }
    \hfill
    \subfloat[Small separation ($\Delta t = 2.6~\mathrm{ns}$)\label{fig:class_count_dt_small}]{
        \includegraphics[width=0.48\textwidth]{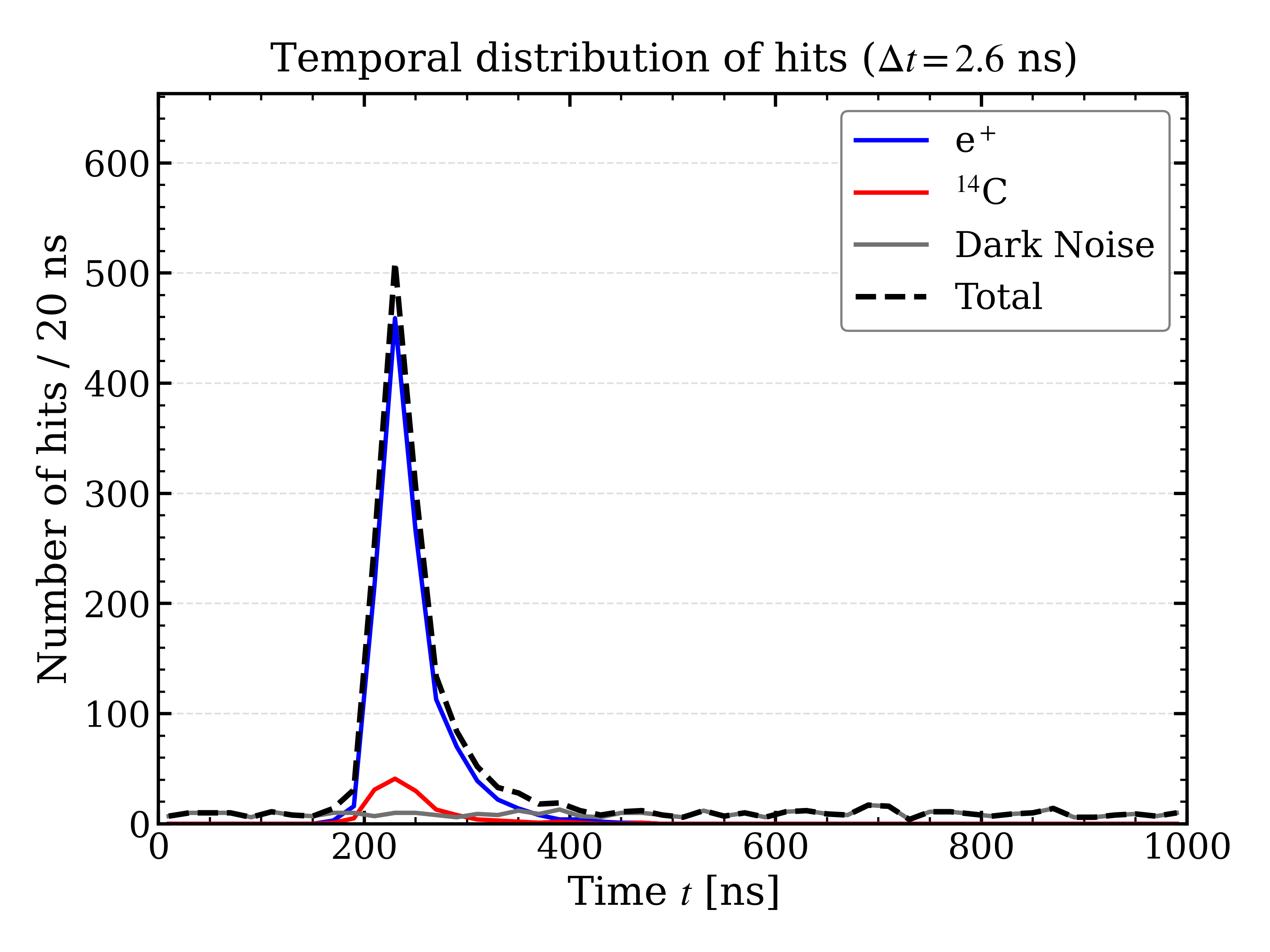}
    }

    \caption{Hit time distributions of pile-up events. The curves show the contributions from $\mathrm{e}^{+}$ annihilation (blue), $^{14}\mathrm{C}$ decay (red), and dark noise (gray), along with the total hit count (dashed black). (a) With large $\Delta t$, the $^{14}\mathrm{C}$ signal appears as a weak secondary peak. (b) With small $\Delta t$, the $^{14}\mathrm{C}$ hits are buried within the dominant $\mathrm{e}^{+}$ peak.}
    \label{fig:time_profiles_comparison}
\end{figure*}

\begin{figure*}[t]
    \centering
    
    \subfloat{%
        \begin{minipage}[b]{0.32\textwidth}
            \centering
            \includegraphics[width=\linewidth]{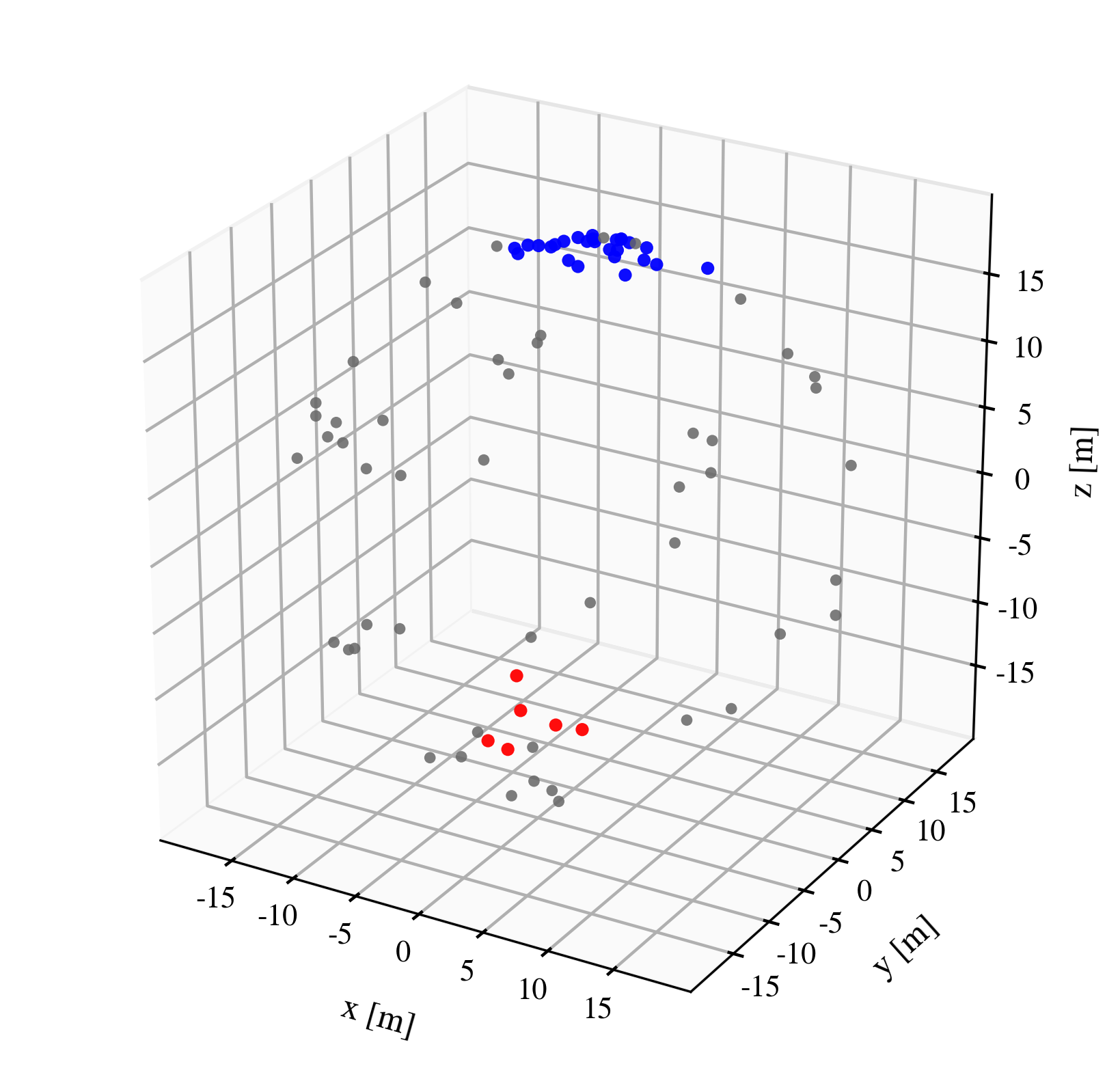}
            
            \vspace{3pt}
            100 -- 200 ns
            \label{fig:t1}
        \end{minipage}%
    }\hfill
    \subfloat{%
        \begin{minipage}[b]{0.32\textwidth}
            \centering
            \includegraphics[width=\linewidth]{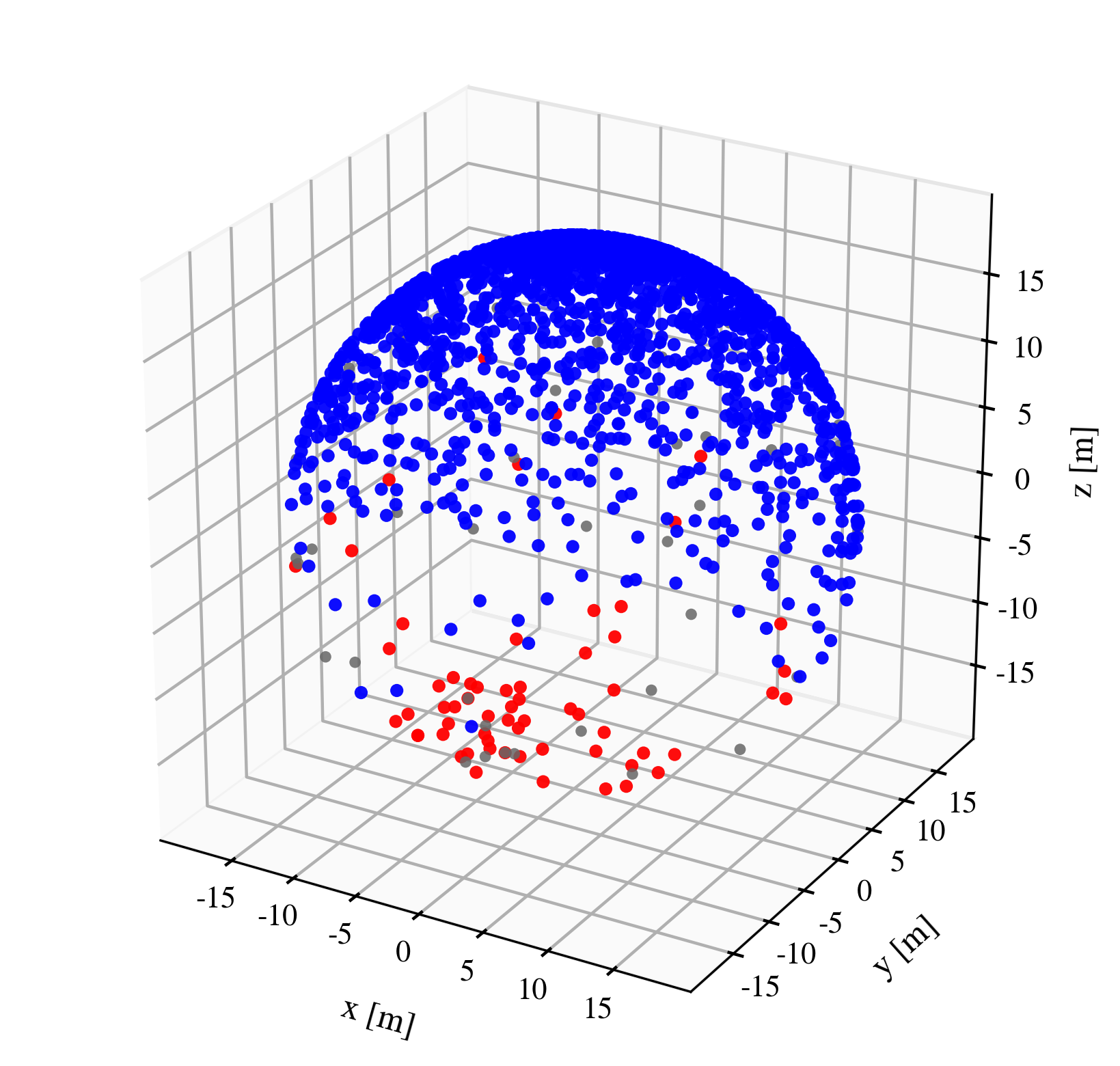}
            
            \vspace{3pt}
            200 -- 300 ns
            \label{fig:t2}
        \end{minipage}%
    }\hfill
    \subfloat{%
        \begin{minipage}[b]{0.32\textwidth}
            \centering
            \includegraphics[width=\linewidth]{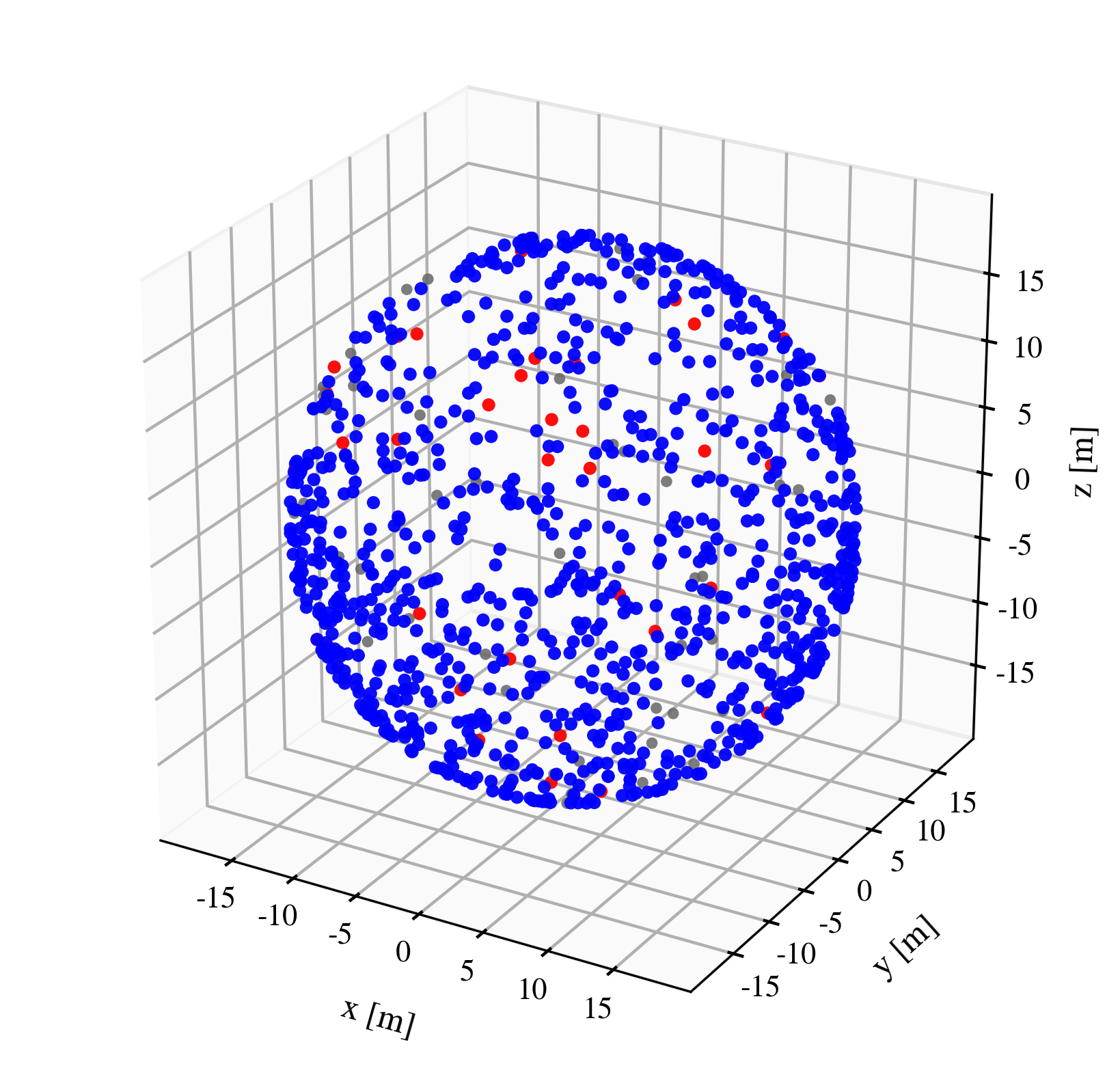}
            
            \vspace{3pt}
            300 -- 400 ns
            \label{fig:t3}
        \end{minipage}%
    }
    
    \vspace{0.5cm}
    
    \subfloat{%
        \begin{minipage}[b]{0.32\textwidth}
            \centering
            \includegraphics[width=\linewidth]{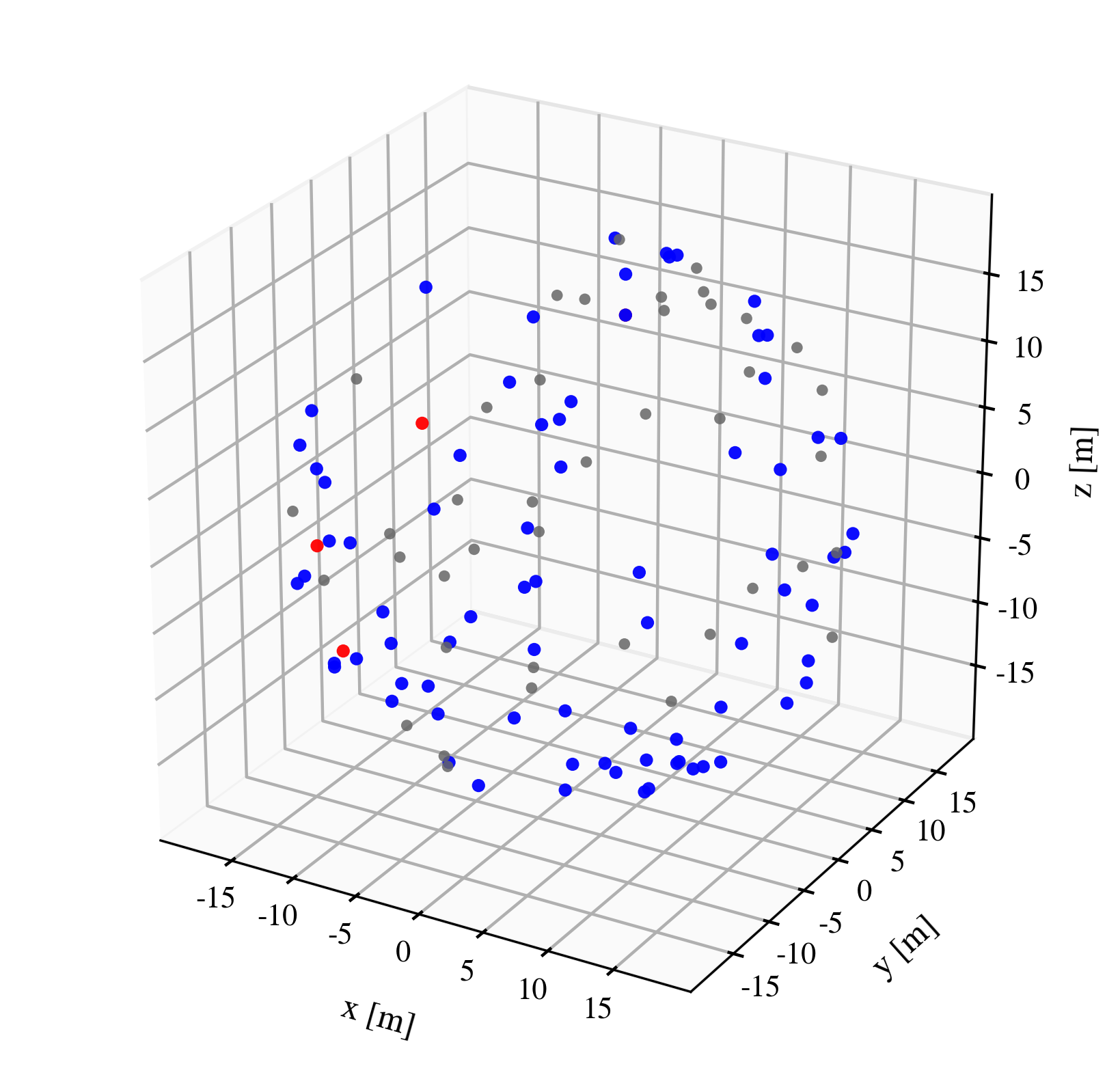}
            
            \vspace{3pt}
            400 -- 500 ns
            \label{fig:t4}
        \end{minipage}%
    }\hfill
    \subfloat{%
        \begin{minipage}[b]{0.32\textwidth}
            \centering
            \includegraphics[width=\linewidth]{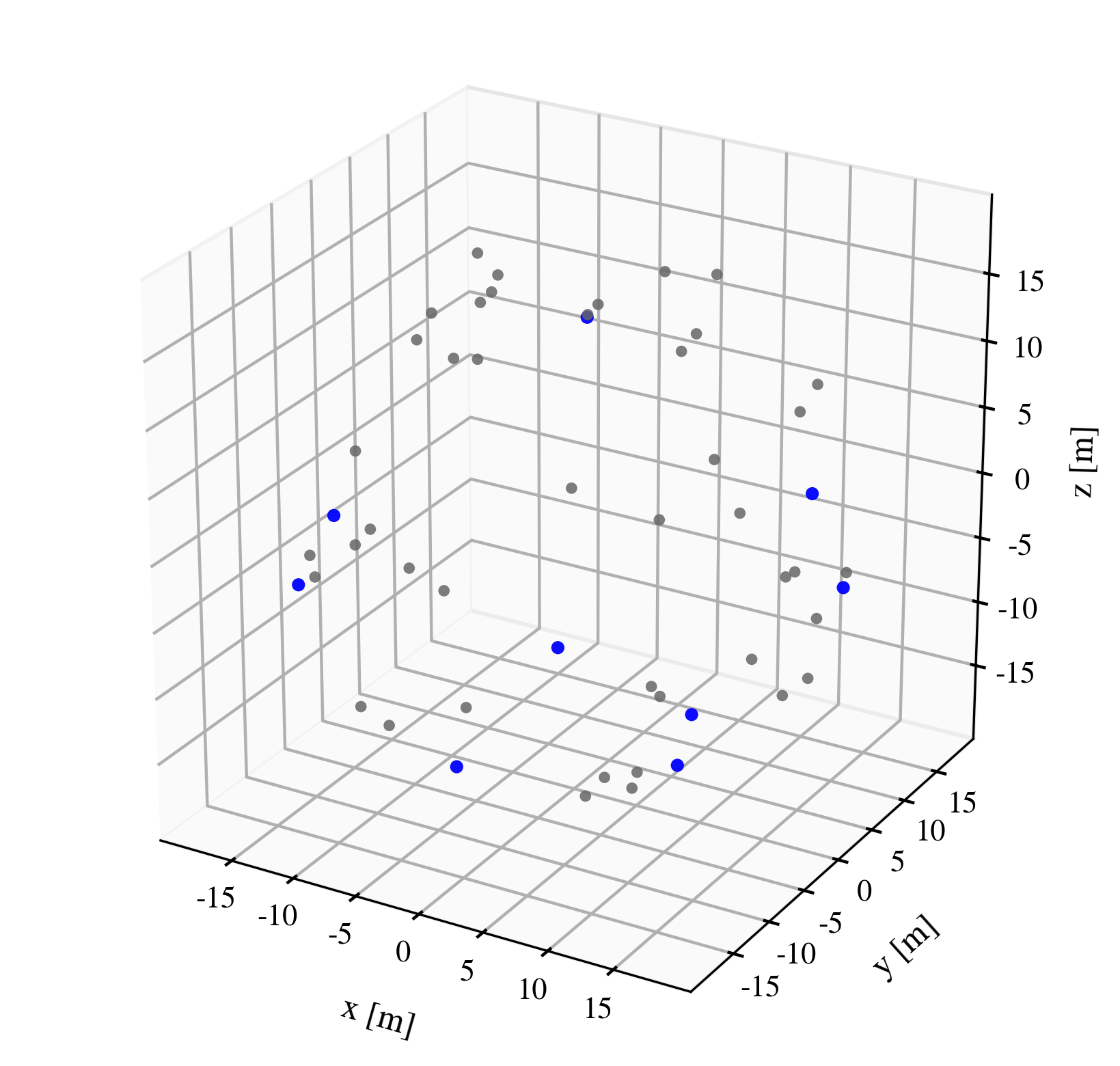}
            
            \vspace{3pt}
            500 -- 600 ns
            \label{fig:t5}
        \end{minipage}%
    }\hfill
    \subfloat{%
        \begin{minipage}[b]{0.32\textwidth}
            \centering
            \includegraphics[width=\linewidth]{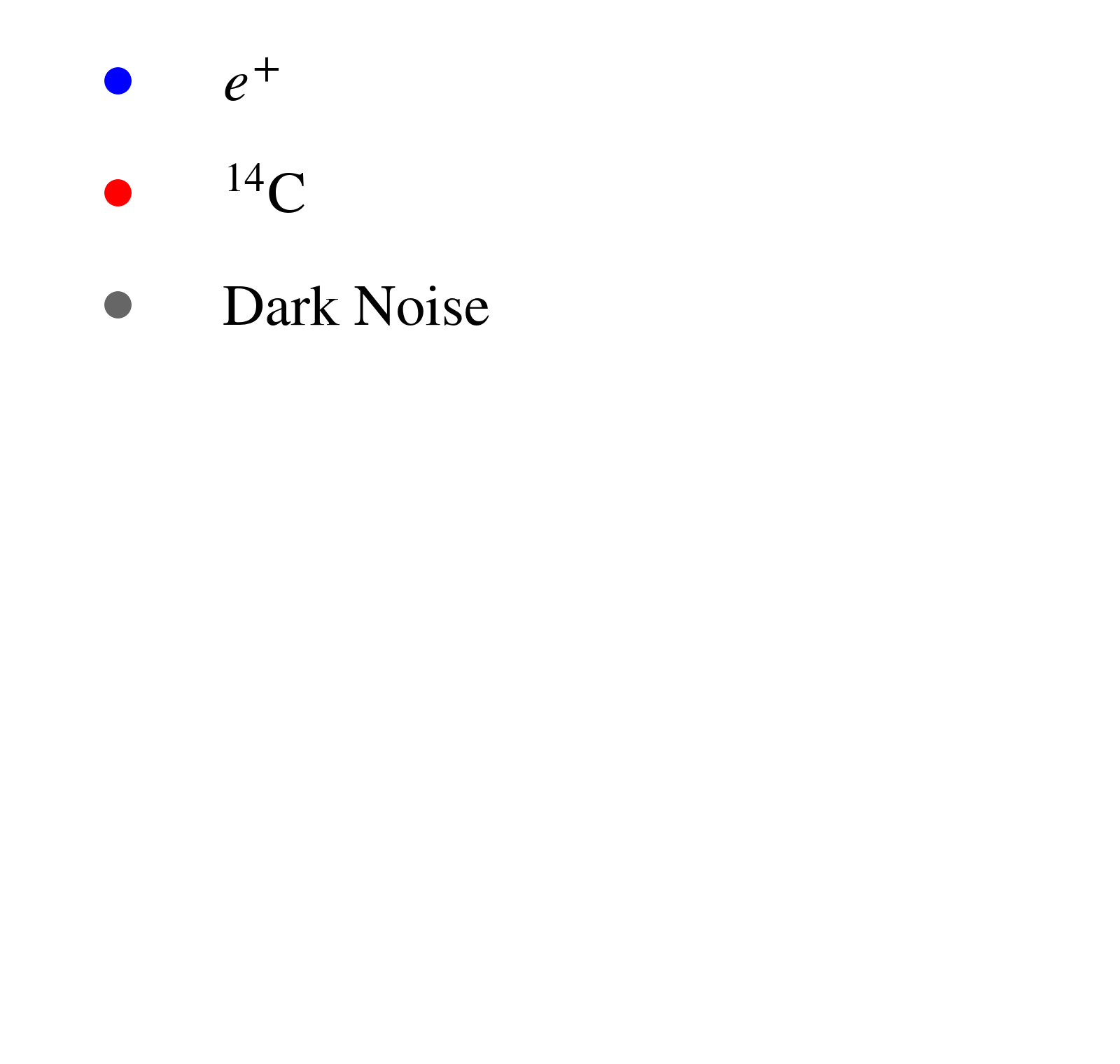}
            \label{fig:info}
        \end{minipage}%
    }
    
    \caption{The spatiotemporal structure of a pile-up event with $\Delta t = 1.9~\mathrm{ns}$. The truth information  for the $\mathrm{e}^{+}$ event consists of a kinetic energy $E_k({\mathrm{e}^{+}})=1.4~\mathrm{MeV}$ and a vertex position of $(x, y, z) = (-1.9, 5.7, 11.1)~\mathrm{m}$, while that for the $^{14}\mathrm{C}$ event consists of a deposited energy $E_{^{14}\mathrm{C}} = 107.4~\mathrm{keV}$ and a vertex position of $(x, y, z) = (-5.4, 1.9, -14.8)~\mathrm{m}$. The plots show the spatial distribution of photon hits in $100~\mathrm{ns}$ intervals. Blue dots represent hits from the positron annihilation, red dots from the $^{14}\mathrm{C}$ decay, and gray dots indicate dark noise.}
    \label{fig:event_evolution}
\end{figure*}

The identification of $^{14}\mathrm{C}$ photon hits can be viewed as a semantic segmentation task on a sparse point cloud defined on a spherical detector geometry. 
Compared to point cloud segmentation problems, this task is complicated by 
the sparsity of the data and the discontinuous distribution of photon hits in both space and time.
To address these challenges, we investigate two classes of deep learning architectures: 
a graph-based model implemented as a Gated Spatiotemporal Graph Neural Network (Gated-STGNN)\cite{GGNN}, and two Transformer-based models~\cite{Vaswani:2017lxt}: the Spatiotemporal Transformer with Scalar Charge Encoding (STT-Scalar) , which uses the charge of each PMT hit directly, and the Spatiotemporal Transformer with Vector Charge Encoding (STT-Vector), which incorporates charge information aggregated from neighboring hits. 
The graph-based approach emphasizes local spatiotemporal correlations through structured neighborhood aggregation, 
while the Transformer framework captures global dependencies via self-attention mechanisms. Their performance is evaluated based on hit identification efficiency and the resulting impact on the reconstructed energy resolution.

The paper is organized as follows. Section~\ref{sec:dataset} describes the simulation and data representation. Section~\ref{sec:method} details the proposed graph and attention-based architectures. The training protocol and evaluation metrics are presented in Section~\ref{sec:train_eval}. Finally, Section~\ref{sec:results} discusses the classification performance and its impact on energy resolution, followed by conclusions in Section~\ref{sec:conclusion}.

\section{Simulation and Dataset}
\label{sec:dataset}

In this work, the training and performance evaluation are based on 
simulated samples. These samples are generated using a detector configuration
similar to the JUNO central detector~\cite{JUNO:2021vlw}, including PMT positions, LS properties, etc.
Various electronic effects, such as PMT transit-time smearing, PMT dark noise and PMT single-photoelectron charge smearing, were implemented in a toy electronics simulation. 
Moreover, as mentioned in the Introduction, the simulation was customized such that one $^{14}\mathrm{C}$ pile-up was enforced for every e$^+$ event.
With a light yield of $1665$ photon hits per MeV from a previous study~\cite{JUNO:2024fdc}, an $\mathrm{e}^{+}$ at rest (\qty{0}{MeV} kinetic energy) deposits around \qty{1.022}~{MeV} energy in the LS and induces approximately 1702 photon hits recorded by the PMTs, with a statistical fluctuation of 41 photons. 
By contrast, a $^{14}\mathrm{C}$ decay produces around 100 photons on average, which are easily overwhelmed by the $\mathrm{e}^{+}$ photon hits.
In addition, given an average rate of 22.7~kHz and 17,612~PMTs, PMT dark noise contributes about 400 photon hits within the acquisition window of each event, making the identification of $^{14}\mathrm{C}$ photons more challenging.

The acquisition window of an event is \qty{1000}{ns}. Each event contains photon hits induced by three components:  one IBD positron ($\mathrm{e}^{+}$),  one $^{14}\mathrm{C}$ $\beta$ decay, and uniformly distributed dark noise (DN). 
For the $i$-th hit, the data vector is defined as
\begin{equation}
\mathbf{s}_i = (x_i, y_i, z_i, t_i, q_i, \ell_i),
\label{eq:hit_vector}
\end{equation}
where $(x, y, z)$ are the spatial coordinates of the triggered PMT~\cite{JUNO:2021vlw}, $t$ is the arrival time relative to the start of the acquisition window, 
and $q$ is the charge. 
Here $\ell_i \in \{0, 1, 2\}$ denotes the truth label of the hit, representing it originating from DN, $\mathrm{e}^{+}$, and $^{14}\mathrm{C}$, respectively.

The photon hit time distributions of two example events are shown in Fig.~\ref{fig:time_profiles_comparison}. The difficulty of hit-level discrimination depends strongly on the relative difference between the $^{14}\mathrm{C}$ decay time ($t_{^{14}\mathrm{C}}$) and the $\mathrm{e}^{+}$ time ($t_{\mathrm{e}^{+}}$), defined as
\begin{equation}
\Delta t \equiv t_{^{14}\mathrm{C}} - t_{\mathrm{e}^{+}}.
\label{eq:delta_t}
\end{equation}
For an event with large $|\Delta t|$, as shown in Fig.~\ref{fig:time_profiles_comparison}(a), the two components are temporally separated and background $^{14}\mathrm{C}$ hits can be rejected using simple timing cuts~\cite{JUNO:2024fdc}. For events with $\Delta t \in [-100, 300]\,\mathrm{ns}$, the $^{14}\mathrm{C}$ component is embedded within the dominant $\mathrm{e}^{+}$ peak, rendering timing information alone insufficient, as indicated in Fig.~\ref{fig:time_profiles_comparison}(b).
A joint analysis of spatial and temporal correlations is therefore required. 
The spatiotemporal structure of a typical event in this regime is shown in Fig.~\ref{fig:event_evolution}. 
This study focuses on events with $\Delta t \in [-100, 300]\,\mathrm{ns}$, excluding other events.

Events at a few-MeV scale contain relatively few photon hits. Thus, only a small fraction of all PMTs are triggered and register photon hits. Moreover, the majority of these PMTs will detect no more than 4 photons in the entire 1000~ns acquisition window.
Therefore, the data structure manifests inherent sparsity and spatiotemporal discontinuity. This is very different from the scenario of conventional point cloud tasks~\cite{qi2017pointnet, fan2021p4transformer, guo2021deep, liu2025mamba4d, lv2025adapting}. 
The samples used in this study are as follows.

\textbf{Training sample:} Models are trained on a sample of one million simulated events with continuous positron kinetic energy $E_k(\mathrm{e^+}) \in [0, 5]~\mathrm{MeV}$ and vertices distributed uniformly within the LS

\textbf{Test sample:} Evaluation is conducted on 24 independent test sets, each containing $8 \times 10^{4}$ events. These data sets correspond to the combinations of six discrete positron kinetic energies,
\begin{equation}
E_k(\mathrm{e}^{+}) \in \{0, 1, 2, 3, 4, 5\}~\mathrm{MeV},
\end{equation}
and four exemplary vertex positions along the detector $z$-axis:
\begin{equation}
z \in \{0,\, 6,\, 15,\, 16.5\}~\mathrm{m}.
\end{equation}

The datasets at 0 and 6 m characterize model performance in the detector’s inner volume. To probe performance near the detector boundary, we also consider datasets at larger radii. In particular, the JUNO central detector exhibits a total reflection region, approximately defined by r $\in [15.6, 17.7] \,\mathrm{m}$~\cite{JUNO:2021vlw}. The dataset at 15 m probes events close to the boundary of this region, while the dataset at 16.5 m corresponds to events well within the total reflection regime and serves as a representative sample.

\section{Methodology}
\label{sec:method}

\subsection{Gated Spatiotemporal Graph Neural Network (Gated-STGNN)}
\label{sec:model_gated_stgnn}

LHC uses Gated-GNN~\cite{GGNN} to suppress the pile-up effect in the jet reconstruction~\cite{ArjonaMartinez:2018eah,PuppiML:Li:2022omf, Shlomi:2020gdn}. In this work, we adapt the Gated-GNN to the $^{14}\mathrm{C}$ hit-level tagging task and refer to it as Gated-STGNN. Gated-STGNN models events as graphs where nodes represent hits and edges encode local spatiotemporal proximity. The architecture employs message passing for contextual propagation and a per-hit classifier for the three target classes. The Gated-STGNN serves as the baseline model for this study.

\paragraph*{Input representation and batching.}

For an event with $N$ hits, the input feature vector $\mathbf{u}_i$ is constructed from the hit vector $\mathbf{s}_i$ defined in Eq.~(\ref{eq:hit_vector}).

The spatial coordinates $\mathbf{r}_i = (x_i, y_i, z_i)$ are mapped to the unit direction vector $\hat{\mathbf{x}}_i = \mathbf{r}_i / \lVert \mathbf{r}_i \rVert$. 

The hit time $t_i$ is transformed into a normalized continuous time $\tilde{t}_i \in [0,1)$ within the acquisition window $[t_{\min}, t_{\max})$ to preserve the precise timing required for microscopic hit-level correlations. To simultaneously capture the macroscopic temporal evolution of the event, $t_i$ is also mapped to a normalized coarse-grained time-bin index $\tilde{b}_i \in [0,1]$ using a bin width $\Delta t_{\mathrm{bin}}$.
The charge $q_i$ is used directly.
The resulting input vector for the $i$-th hit is given by
\begin{equation}
    \mathbf{u}_i = (\hat{\mathbf{x}}_i, q_i, \tilde{t}_i, \tilde{b}_i) \in \mathbb{R}^{6}.
    \label{eq:input_vector}
\end{equation}
The number of hits varies across events; however, parallel processing on GPUs requires a uniform sequence length for all events within a mini-batch. Events are therefore zero-padded so that the number of hits in every event equals $L_{\max} = \max_m N_m$, where $N_m$ denotes the valid hit number of the $m$-th event in the batch. Since zero-padding can interfere with the hit identification task, a binary mask $M$ is employed to strictly exclude padded entries from the graph construction and message passing operations:

\begin{equation}
    M_{mj} = \begin{cases} 
    1, & \text{if } j \ge N_m \quad \text{(padding)}, \\ 
    0, & \text{otherwise} \quad \text{(valid)}, 
    \end{cases}
    \label{eq:padding_mask}
\end{equation}
where the index $m$ denotes the event within the mini-batch, and $j \in [0, L_{\max}-1]$ represents the hit index within that specific event.

\paragraph*{Window-level context.}

To capture the global temporal evolution, aggregate statistics—specifically the hit number, the sum and mean of the charges, alongside the mean and variance of the unit direction vectors—are computed for each discrete time bin $b$ to serve as the bin feature vector $\mathbf{w}_b$. The sequence of bin features $\mathbf{w}_b$ is then processed by a Gated Recurrent Unit (GRU) to generate hidden states $\mathbf{c}_b$~\cite{Cho:2014sdq}. Each hit $i$ is subsequently assigned the state $\mathbf{c}_{b_i}$ corresponding to its specific time bin, where the integer bin index is given by $b_i = \lfloor (t_i - t_{\min}) / \Delta t_{\mathrm{bin}} \rfloor$. The initial hit embedding is then obtained by applying an input multilayer perceptron (MLP) to the concatenated hit and context features: $\mathbf{h}_i^{(0)} = \mathrm{MLP}_{\mathrm{in}}([\mathbf{u}_i; \mathbf{c}_{b_i}])$.

\paragraph*{Graph construction and edge features.}
For each hit $i$, a localized spatiotemporal neighborhood is strictly defined as the set $\mathcal{N}_i = \{ j \neq i \mid |t_j-t_i| \le \Delta t_{\mathrm{nb}}, \lVert \mathbf{x}_j-\mathbf{x}_i \rVert \le \Delta r_{\mathrm{nb}} \}$. To maintain a fixed graph degree for batched processing, a multi-set $\mathcal{S}_i$ of exactly $K$ indices is constructed by sampling uniformly without replacement from $\mathcal{N}_i$. If $|\mathcal{N}_i| < K$, the sequence is padded with self-loops ($j=i$). The edge features $\mathbf{e}_{ij}$ are computed for each $j \in \mathcal{S}_i$ and encompass the Fourier-encoded absolute time difference, the angular alignment $\cos\theta_{ij}$, the normalized spatial separation, the charge difference, and the normalized signed time difference.

\paragraph*{Gated message passing.}
The model applies $L$ message-passing layers. At layer $\ell$, messages are computed via an MLP on the neighbor states alongside the corresponding edge features, and then aggregated by a mean operator over the $K$ sampled nodes: $\bar{\mathbf{m}}_i^{(\ell)} = \frac{1}{K}\sum_{j \in \mathcal{S}_i} \mathrm{MLP}_{\mathrm{msg}}^{(\ell)}([\mathbf{h}_j^{(\ell)}; \mathbf{e}_{ij}])$. The node state is subsequently updated using a GRU cell: $\mathbf{h}_i^{(\ell+1)} = \mathrm{GRUCell}(\bar{\mathbf{m}}_i^{(\ell)}, \mathbf{h}_i^{(\ell)})$.

\paragraph*{Output and implementation.}
The final latent representations $\mathbf{h}_i^{(L)} \in \mathbb{R}^H$, where $H$ denotes the hidden feature dimension, are projected onto the decision space via a position-wise output MLP, yielding a vector of class logits $\mathbf{s}_i \in \mathbb{R}^{3}$. The three components of $\mathbf{s}_i$ correspond to the confidence scores for the dark noise, positron ($\mathrm{e}^{+}$), and $^{14}\mathrm{C}$ categories, respectively. The baseline configuration utilizes $H=128$, $L=3$ message-passing layers, a neighborhood size of $K=24$, spatial and temporal cutoffs of $\Delta r_{\mathrm{nb}}=9000~\mathrm{mm}$ and $\Delta t_{\mathrm{nb}}=10~\mathrm{ns}$, and a bin width of $\Delta t_{\mathrm{bin}} = 20~\mathrm{ns}$ over the acquisition window $[0, 1000)~\mathrm{ns}$. This yields a compact model with approximately 0.66 million trainable parameters. Training is performed using the AdamW optimizer~\cite{Loshchilov:2017bsp} with cosine annealing and mixed-precision arithmetic.

\subsection{Spatiotemporal Transformer with Scalar Charge Encoding (STT-Scalar)}
\label{sec:model_stt_scalar}

The STT-Scalar treats hits as tokens in a sequence, utilizing global self-attention to facilitate information exchange across hits for per-hit classification.

\paragraph*{Input representation.}
For batch size $B$, the input tensor $\mathbf{X} \in \mathbb{R}^{B \times L_{\max} \times 5}$ comprises observables $(x, y, z, t, q)$ from Eq.~\eqref{eq:hit_vector}. Similar to the Gated-STGNN, sequences are padded to $L_{\max}$ and processed using the binary mask $M$ defined in Eq.~\eqref{eq:padding_mask} to prevent padded tokens from attending to valid signals in the self-attention mechanism.

\paragraph*{Spatiotemporal feature encoding.}
Position and time are normalized by $S_x = 10^{4}~\mathrm{mm}$ and $S_t = 10^{3}~\mathrm{ns}$, respectively, and mapped using periodic encodings $\psi(\cdot)$ with $n_{\mathrm{pe}}=5$:
\begin{equation}
\psi(u) = \Big( u, \; \{ \sin(k\pi u),\; \cos(k\pi u) \}_{k=1}^{n_{\mathrm{pe}}} \Big).
\label{eq:stt_scalar_pe}
\end{equation}
The scalar charge $q$ is embedded via an MLP to $\mathbf{e}_q \in \mathbb{R}^{32}$. These features are concatenated and projected to dimension $d_{\mathrm{model}}$:
\begin{equation}
\mathbf{h}^{(0)} = \mathrm{Proj}\left(\Big[\psi(\tilde{\mathbf{x}});\; \psi(\tilde{t});\; \mathbf{e}_q\Big]\right) \in \mathbb{R}^{d_{\mathrm{model}}}.
\end{equation}

\paragraph*{Transformer backbone and output.}
A stack of $L$ pre-norm Transformer encoder layers $\mathcal{T}^{(\ell)}$ processes the embeddings:
\begin{equation}
\mathbf{h}^{(\ell+1)} = \mathcal{T}^{(\ell)}\!\left(\mathbf{h}^{(\ell)}\right), \qquad \ell=0,\ldots,L-1.
\end{equation}
The final hidden representation of each hit is passed through a multilayer perceptron to produce a vector of classification logits $\mathbf{s}_i \in \mathbb{R}^{3}$, corresponding to the three hit categories: dark noise, positron-induced signal, and $^{14}\mathrm{C}$ background. 

\paragraph*{Implementation details.}
The architecture uses $L=6$, $d_{\mathrm{model}}=512$, and $n_{\mathrm{head}}=4$, totaling approximately 19.09 million trainable parameters. Optimization employs AdamW with a cosine-annealing schedule.

\subsection{Spatiotemporal Transformer with Vector Charge Encoding (STT-Vector)}
\label{sec:model_stt_vector}

\begin{figure*}[htbp]
  \centering
  \includegraphics[width=0.9\textwidth]{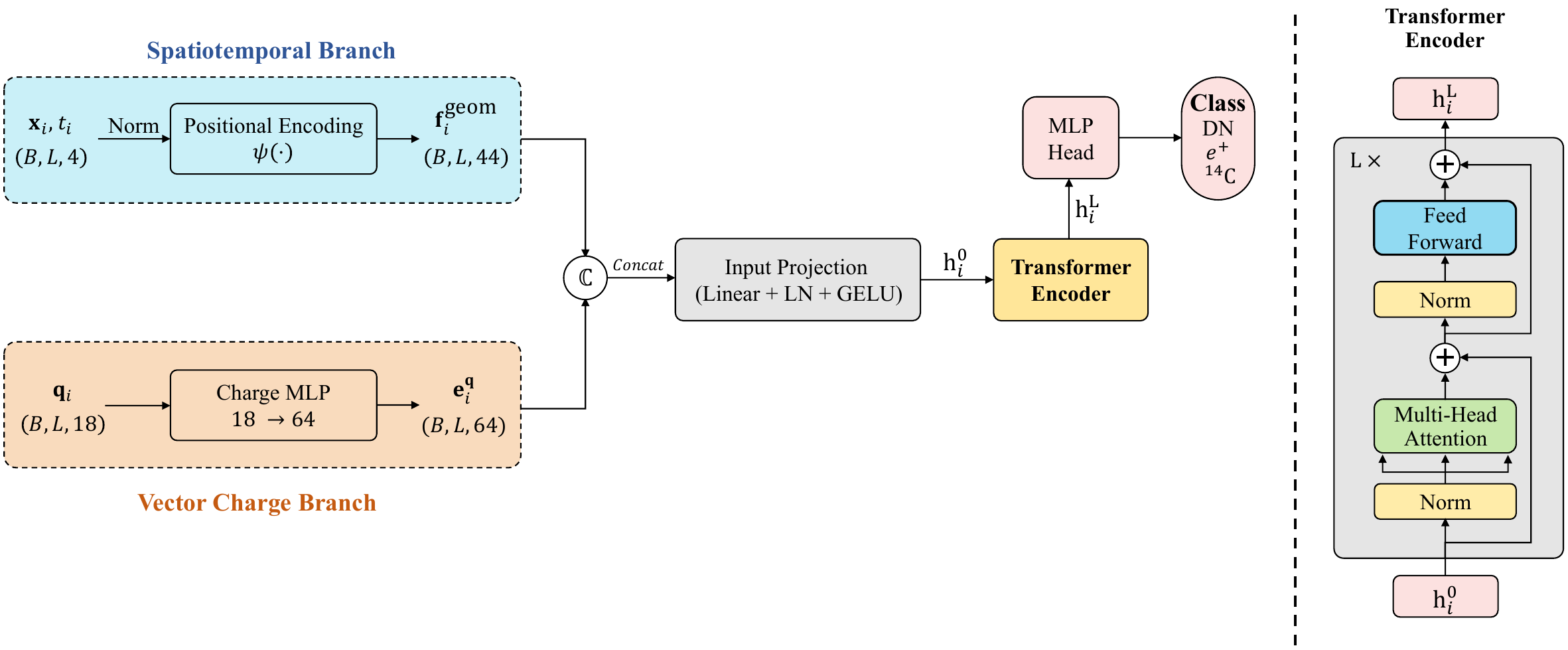}
  \caption{
    Architecture of STT-Vector. 
    Each hit is processed through two parallel encoding branches.
    The spatiotemporal branch (top) maps normalized spatial coordinates and hit times to a high-dimensional representation using sinusoidal positional encodings $\psi(\cdot)$.
    The vector charge branch (bottom) embeds the 18-dimensional charge feature vector $\mathbf{q}_i$, which summarizes multi-scale local and global charge accumulation, using a dedicated multilayer perceptron.
    The two embeddings are concatenated, projected to the Transformer model dimension $d_{\mathrm{model}}$, and passed through $L$ Transformer encoder layers to produce per-hit class logits for dark noise, $\mathrm{e}^{+}$, and $^{14}\mathrm{C}$.
    }
  \label{fig:stt_vector_arch}
\end{figure*}

The STT-Vector model, inspired by token-based Transformer models
such as the Particle Transformer~\cite{Qu:2022mxj}, augments the STT-Scalar by enriching each hit token with an 18-dimensional charge feature vector $\mathbf{q}_i$, explicitly encoding local and global spatiotemporal charge density correlations. The overall architecture is shown in Fig.~\ref{fig:stt_vector_arch}.

\paragraph*{Vector charge encoding.}
For each hit $i$, the vector $\mathbf{q}_i$ is constructed based on charge accumulation within defined temporal frames relative to $t_i$: current ($\mathcal{T}^{(0)}_i: |\Delta t| \le 5\,\mathrm{ns}$), next ($\mathcal{T}^{(+)}_i: 5 < \Delta t \le 15\,\mathrm{ns}$), and previous ($\mathcal{T}^{(-)}_i: -15 \le \Delta t < -5\,\mathrm{ns}$).
We define the charge sum operator $S_i(\tau, R) = \sum_{j \in \mathcal{T}^{(\tau)}_i \cap \mathcal{B}_i(R)} q_j$ over frame $\tau$ and spatial radius $R$, and the temporal asymmetry $\Delta S_i(\tau', R, R') = S_i(0, R) - S_i(\tau', R')$.

The feature vector is structured as:
\begin{equation}
    \mathbf{q}_i = \Big[ q_i; \; \mathbf{v}_{\text{glob}}; \; \mathbf{v}^{(1)}_{\text{loc}}; \; \mathbf{v}^{(2)}_{\text{loc}}; \; \mathbf{v}^{(3)}_{\text{loc}}; \; \mathbf{v}^{(4)}_{\text{loc}} \Big]^\top \in \mathbb{R}^{18}.
\end{equation}
The global block $\mathbf{v}_{\text{glob}}$ captures total event activity ($R \to \infty$):
\begin{equation}
    \mathbf{v}_{\text{glob}, i} = \Big( S_i(0), S_i(+), S_i(-), \Delta S_i(+), \Delta S_i(-) \Big)_{\infty}.
\end{equation}
The four local blocks $\mathbf{v}^{(k)}_{\text{loc}}$ encode density at spatial scales defined by radius pairs $(R_k, R'_k) \in \{(3,5), (5,9), (10,16), (16,23)\}\,\mathrm{m}$:
\begin{equation}
    \mathbf{v}^{(k)}_{\text{loc}} = \Big( S_i(0, R_k), \; S_i(+, R'_k), \; \Delta S_i(+, R_k, R'_k) \Big).
\end{equation}
This vector is precomputed for every hit, resulting in an input tensor $\mathbf{X}\in\mathbb{R}^{B\times L_{\max}\times 22}$.

\paragraph*{Model Architecture.}
Spatiotemporal coordinates $(x,y,z,t)$ are encoded using the periodic function $\psi(\cdot)$ as in STT-Scalar. The charge feature vector $\mathbf{q}_i$ is independently embedded via an MLP to a dimension of $d_q=64$. These spatial and charge embeddings are concatenated and projected to the Transformer model dimension $d_{\mathrm{model}}=512$.
The sequence is processed by $L=6$ Transformer encoder layers ($n_{\mathrm{head}}=4$), resulting in a total of approximately 19.11 million trainable parameters.
The sequence padding, the masking mechanism defined in Eq.~\eqref{eq:padding_mask}, and the final output MLP mapping to the three-class logits $\mathbf{s}_i$ are identical to those employed in the STT-Scalar model.

\subsection{Computational complexity and practical considerations}
\label{sec:complexity_practical}

We analyze the computational complexity and memory footprint of the three models to address feasibility for hit multiplicities $N\sim\mathcal{O}(10^3\text{--}10^4)$. Estimates assume batch size $B$, hidden dimension $H$ (which corresponds to $d_{\mathrm{model}}$ in the STT architectures), neighbor count $K$ (graph models), layer count $L$, and attention heads $n_{\mathrm{head}}$ (Transformers).

\subsubsection{Gated-STGNN: Sparse neighborhood message passing}
\label{sec:complexity_gated_stgnn}

Gated-STGNN utilizes sparse message passing on a graph with bounded degree $K$, yielding an edge set size $E \simeq N K$.

\paragraph*{Computational complexity.}
The dominant costs per layer are MLP-based message computation and GRU state updates. With matrix-vector multiplications scaling quadratically in feature dimension, the per-layer cost is $\mathcal{O}(B N K H^2)$. The total scaling is linear in $N$:
\begin{equation}
\mathcal{C}_{\mathrm{Gated-STGNN}} \sim \mathcal{O}(L \cdot B N K H^2).
\end{equation}

\paragraph*{Memory footprint.}
Training memory is dominated by node states and edge tensors required for backpropagation. The scaling per layer is
\begin{equation}
\mathcal{M}_{\mathrm{layer}} \sim \mathcal{O}( B N K H + B N K d_e ),
\end{equation}
where $d_e$ is the edge-feature dimension.

\paragraph*{Neighbor construction cost.}
Neighbor selection employs vectorized dense operations for GPU throughput, incurring a distance computation cost of $\mathcal{O}(B N^2)$ in both compute and memory. However, because this step involves only low-dimensional geometric distance calculations without dense neural network parameter multiplications, its empirical runtime overhead is negligible compared to the $\mathcal{O}(L \cdot B N K H^2)$ network scaling. This preprocessing step is performed once per event and is independent of the message-passing layers.

\subsubsection{STT-Scalar: Global self-attention with scalar charge}
\label{sec:complexity_stt_scalar}

STT-Scalar processes hits as tokens with global self-attention in each layer.

\paragraph*{Computational complexity.}
The cost is dominated by self-attention matrix computation and value aggregation. The $N^2$ token interactions lead to a total complexity scaling of
\begin{equation}
\mathcal{C}_{\mathrm{STT\text{-}Scalar}} \sim \mathcal{O}\!\left(B\, L\, N^2\, H\right).
\end{equation}
This quadratic dependence on $N$ limits scalability at high multiplicities.

\paragraph*{Memory footprint.}
Training memory is dominated by the storage of attention weights and intermediate activations:
\begin{equation}
\mathcal{M}_{\mathrm{attn}} \sim \mathcal{O}\!\left(B\, L\, n_{\mathrm{head}}\, N^2\right).
\end{equation}
The $N^2$ attention tensor typically constrains the maximum feasible batch size.

\subsubsection{STT-Vector: Global self-attention with vector charge features}
\label{sec:complexity_stt_vector}

STT-Vector shares the Transformer backbone of STT-Scalar but includes an additional feature-construction step.

\paragraph*{Computational complexity (network).}
The network complexity remains identical to STT-Scalar at leading order:
\begin{equation}
\mathcal{C}_{\mathrm{STT\text{-}Vector}}^{\mathrm{net}} \sim \mathcal{O}\!\left(B\, L\, N^2\, H\right).
\end{equation}
The 18-dimensional vector embedding adds a subleading $\mathcal{O}(B N)$ cost.

\paragraph*{Charge-feature construction cost.}
Aggregating charges within spatiotemporal windows involves evaluating hit pairs, scaling as:
\begin{equation}
\mathcal{C}_{\mathrm{feat}} \sim \mathcal{O}\!\left(N^2\right).
\end{equation}
In this study, these features are precomputed to decouple the $\mathcal{O}(N^2)$ preprocessing burden from the training loop.

\paragraph*{Summary.}
\label{sec:complexity_summary}
Although neighbor construction inherently entails an $\mathcal{O}(N^2)$ geometric calculation, the dominant computational cost and bottleneck of the Gated-STGNN are governed by the dense matrix multiplications of the sparse neighborhood aggregations. Consequently, within the empirical regime of typical event sizes, the effective training time of the Gated-STGNN scales linearly with the hit number $N$, following $\mathcal{O}(NK)$. In contrast, the Transformer models exhibit a quadratic $\mathcal{O}(N^2)$ complexity in both computation and memory footprint due to the global self-attention mechanism.
The STT-Vector architecture incurs an additional $\mathcal{O}(N^2)$ preprocessing overhead required for constructing the pairwise vector features. 
Benchmarks on a single NVIDIA A6000 GPU (48~GB VRAM) indicate that the Gated-STGNN requires approximately an order of magnitude less training time per epoch than the STT models. 
This disparity reflects the linear versus quadratic complexity scaling with the hit number $N$, compounded by memory constraints that necessitate a reduced batch size for the Transformers ($B=12$) relative to the Gated-STGNN ($B=16$). 
The per-epoch training run times of STT-Scalar and STT-Vector are comparable, as vector feature construction is offloaded to pre-processing.

\begin{figure*}[t]
    \centering
    \subfloat{%
        \begin{minipage}[b]{0.47\textwidth}
            \centering
            \includegraphics[width=\textwidth]{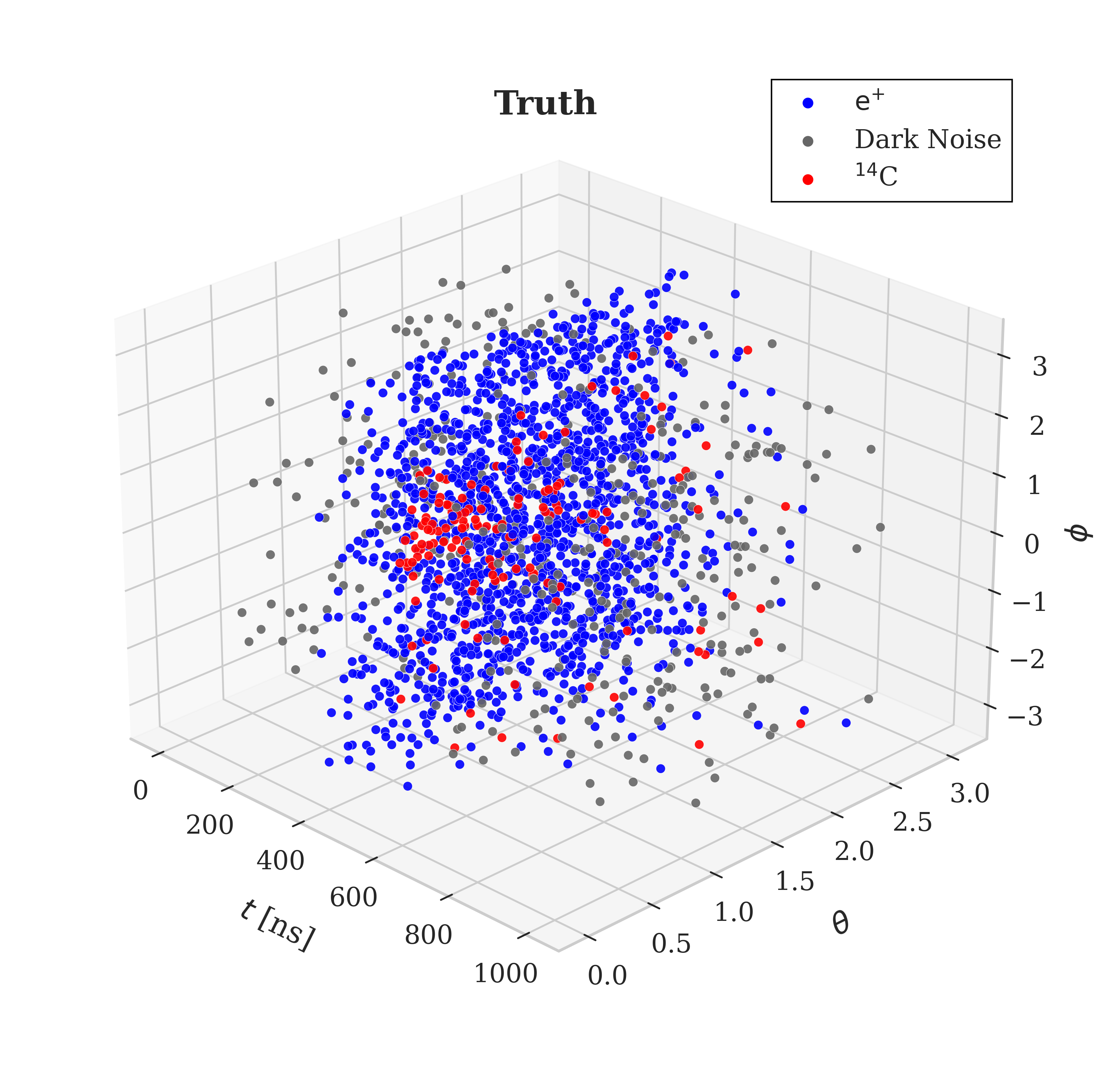}
            
            \vspace{2pt}
            {\small Truth}
            \label{fig:ev_truth}
        \end{minipage}%
    }\hfill
    \subfloat{%
        \begin{minipage}[b]{0.47\textwidth}
            \centering
            \includegraphics[width=\textwidth]{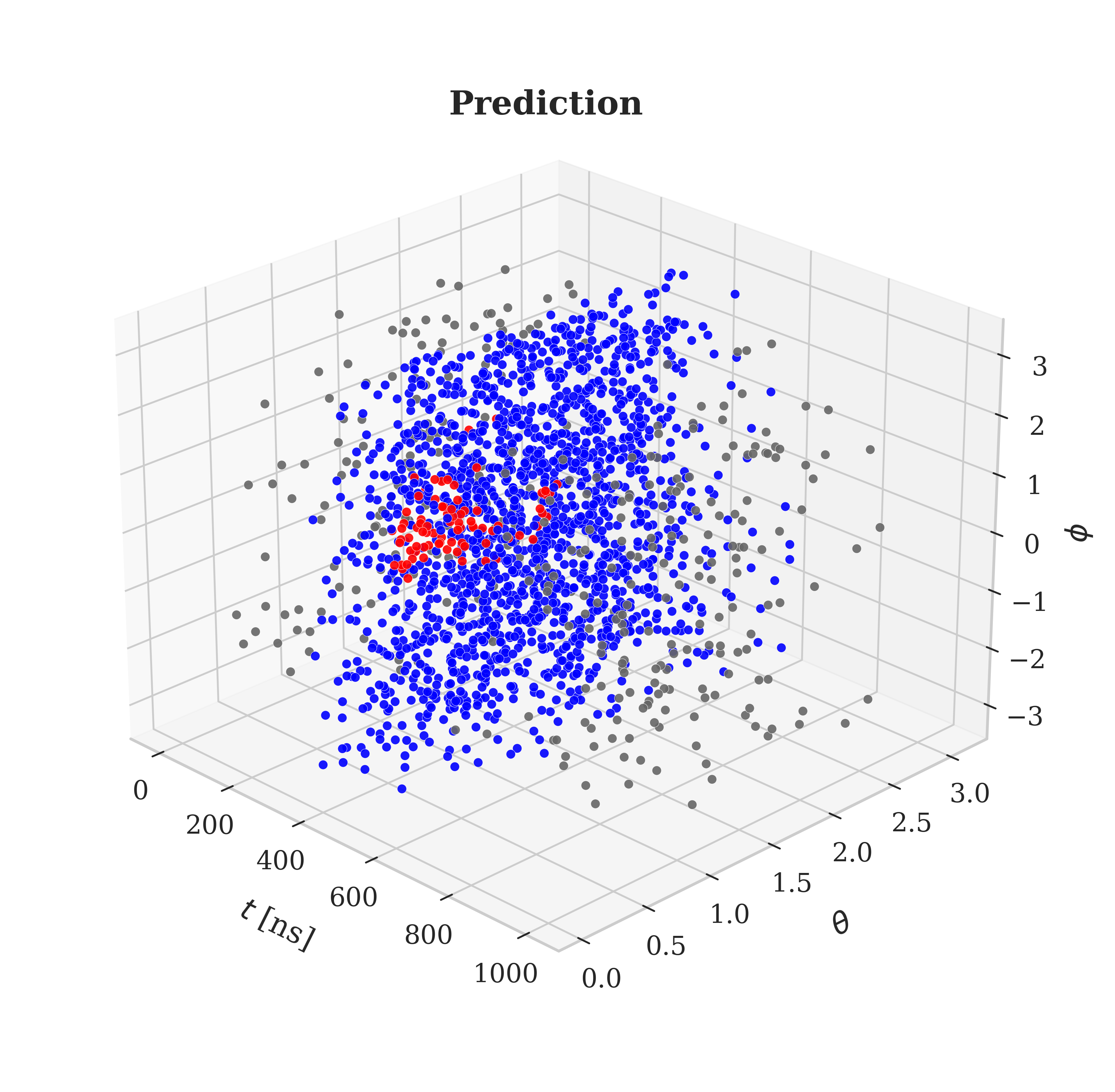}
            
            \vspace{2pt}
            {\small Gated-STGNN}
            \label{fig:ev_stgnn}
        \end{minipage}%
    }

    \vspace{4pt}

    \subfloat{%
        \begin{minipage}[b]{0.47\textwidth}
            \centering
            \includegraphics[width=\textwidth]{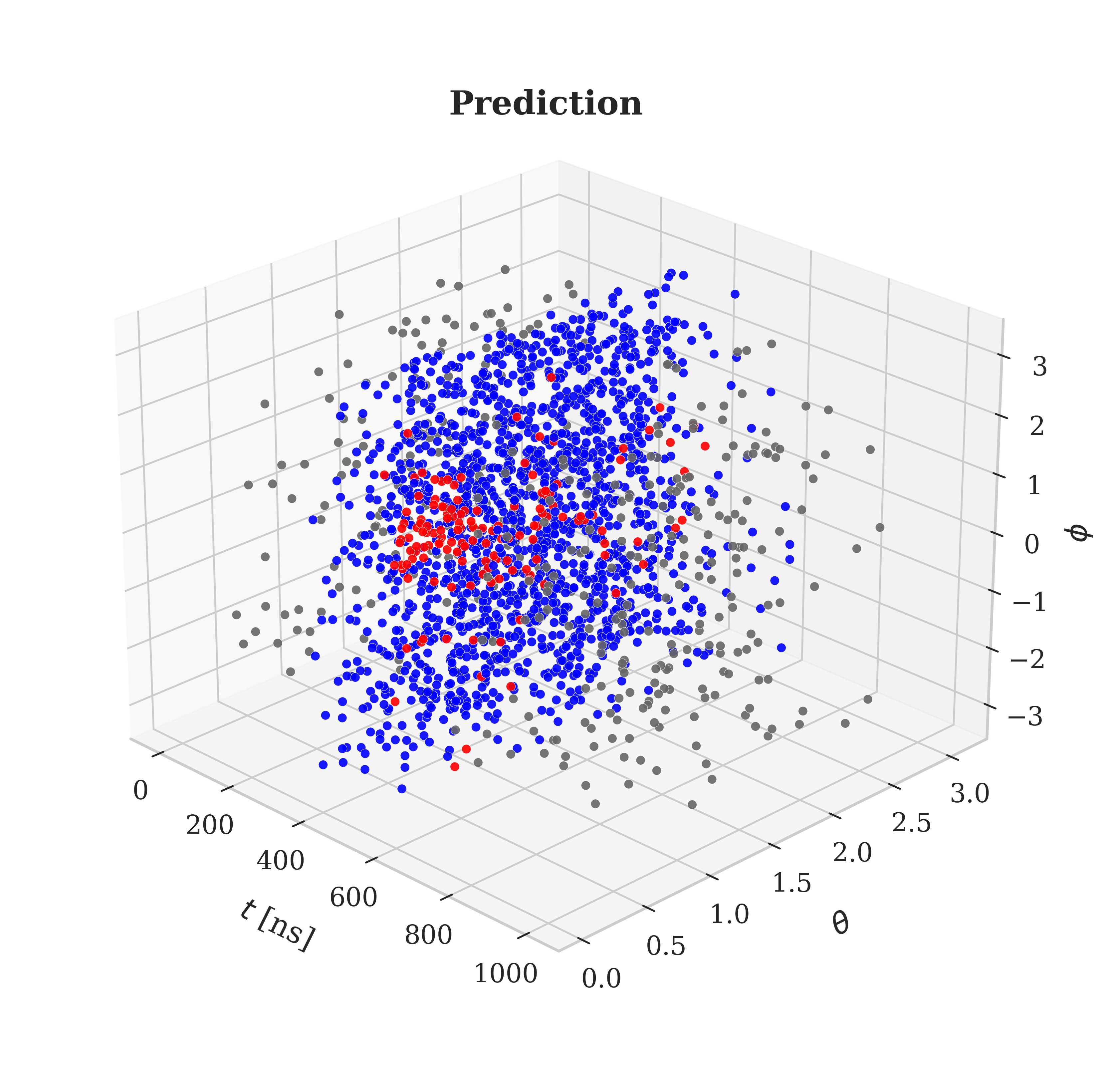}
            
            \vspace{2pt}
            {\small STT-Scalar}
            \label{fig:ev_scalar}
        \end{minipage}%
    }\hfill
    \subfloat{%
        \begin{minipage}[b]{0.47\textwidth}
            \centering
            \includegraphics[width=\textwidth]{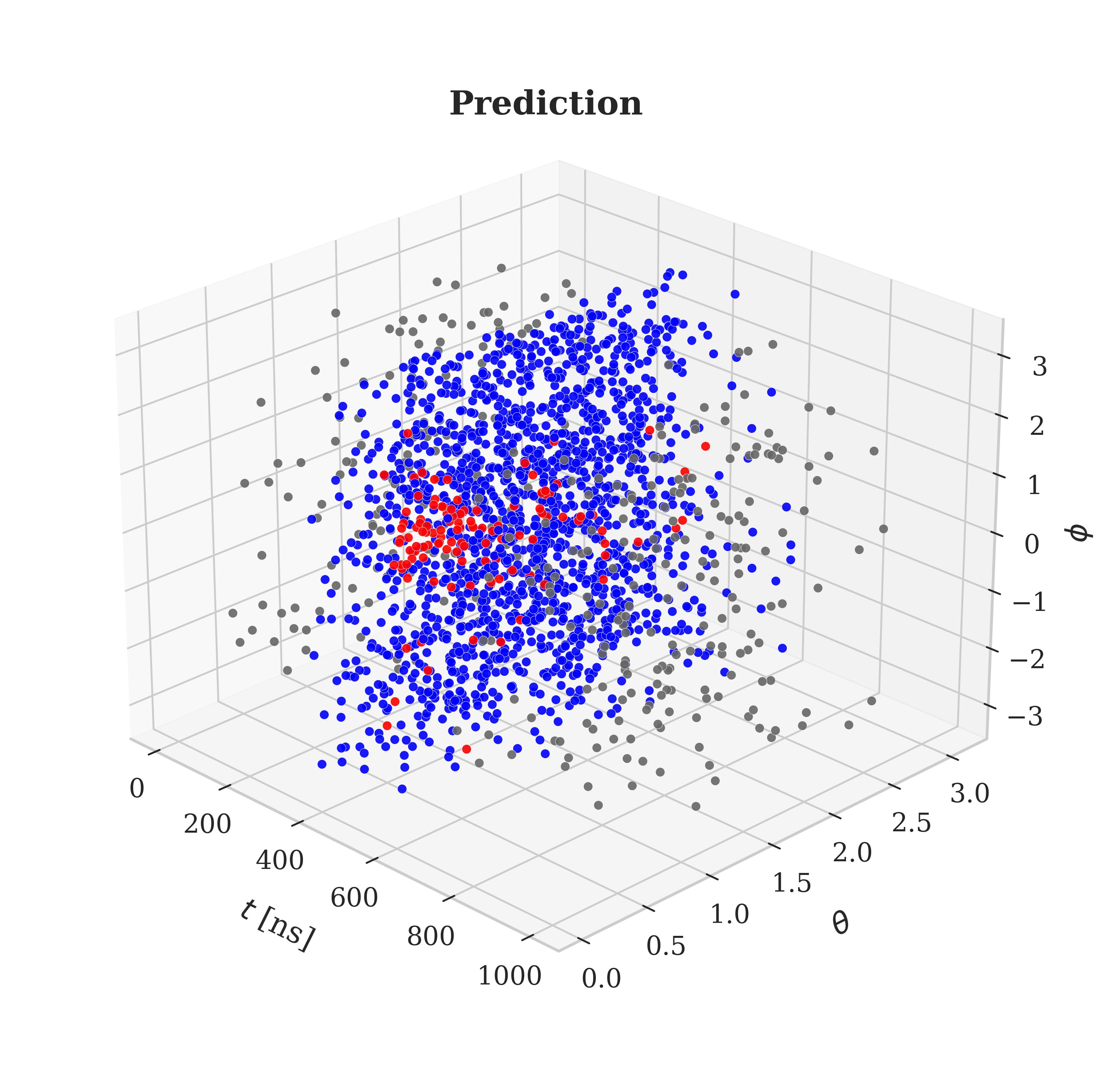}
            
            \vspace{2pt}
            {\small STT-Vector}
            \label{fig:ev_vector}
        \end{minipage}%
    }
    
    \caption{A visualization of the hit identification results of Gated-STGNN, STT-Scalar, and STT-Vector, compared with the truth for a pile-up event with $\Delta t = 165.0~\mathrm{ns}$. The truth information for the $\mathrm{e}^{+}$ event consists of a kinetic energy  $E_k({\mathrm{e}^{+}})=0~\mathrm{MeV}$ and a vertex position of $(x, y, z) = (0.03, -0.02, -0.08)~\mathrm{m}$, while that for the $^{14}\mathrm{C}$ event consists of a deposited energy $E_{^{14}\mathrm{C}} = 82.5~\mathrm{keV}$ and a vertex position of $(x, y, z) = (7.65, 5.75, 11.23)~\mathrm{m}$. The plots display the $(\theta, \phi, t)$ distribution of the hits, where $t$ is the hit time and $(\theta, \phi)$ are the polar and azimuthal angles of the PMT. Blue dots represent hits from the $\mathrm{e}^{+}$ event, red dots from the $^{14}\mathrm{C}$ decay, and gray dots indicate dark noise.}
    \label{fig:3d_event_display}
\end{figure*}

\section{Training and Evaluation}
\label{sec:train_eval}

\begin{figure*}[htbp]
    \centering
    
    \subfloat{%
        \begin{minipage}[b]{0.32\textwidth}
            \centering
            \includegraphics[width=\textwidth]{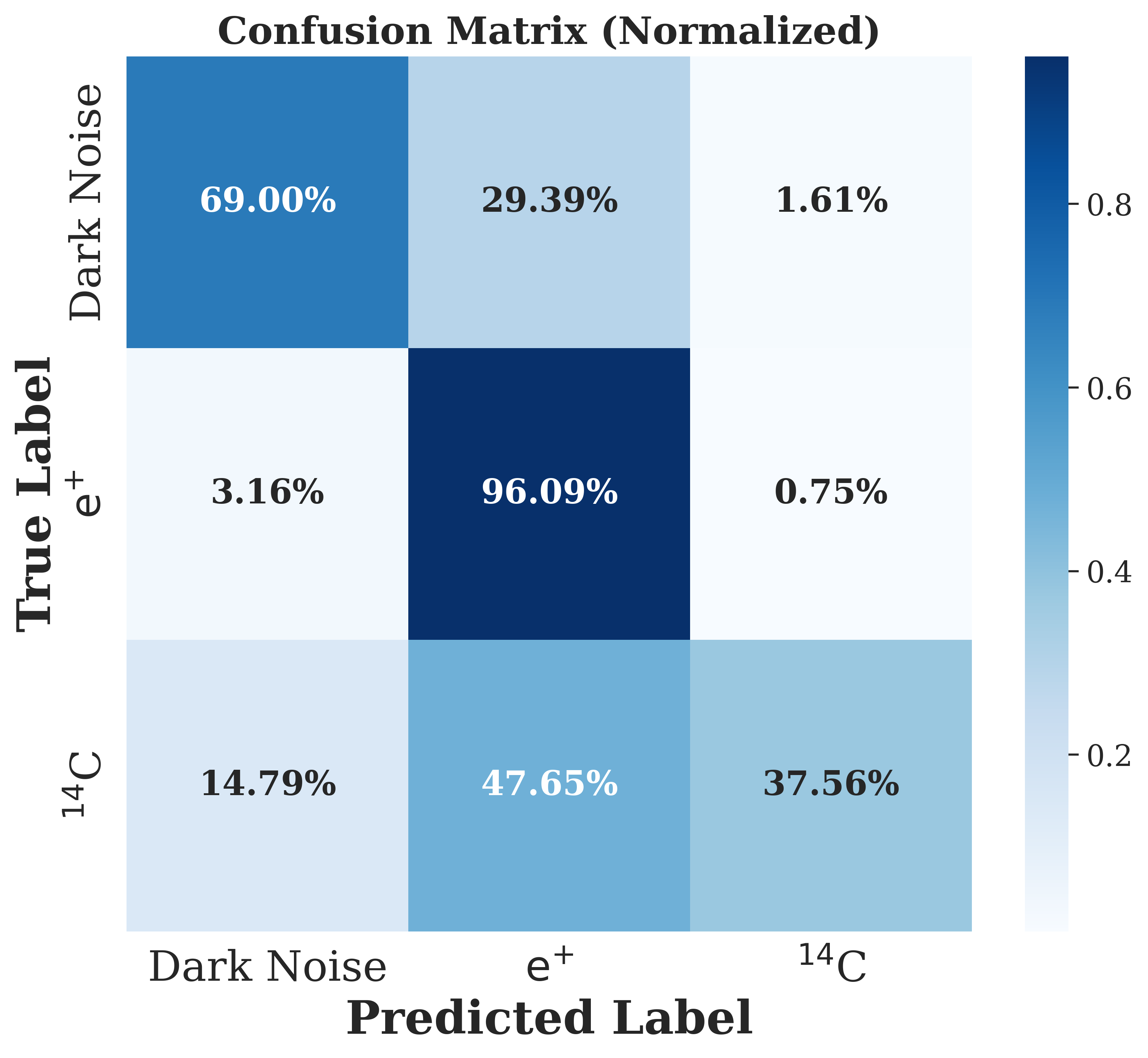}
            
            \vspace{3pt}
            Gated-STGNN
            \label{fig:cm_0MeV_gated}
        \end{minipage}%
    }\hfill
    \subfloat{%
        \begin{minipage}[b]{0.32\textwidth}
            \centering
            \includegraphics[width=\textwidth]{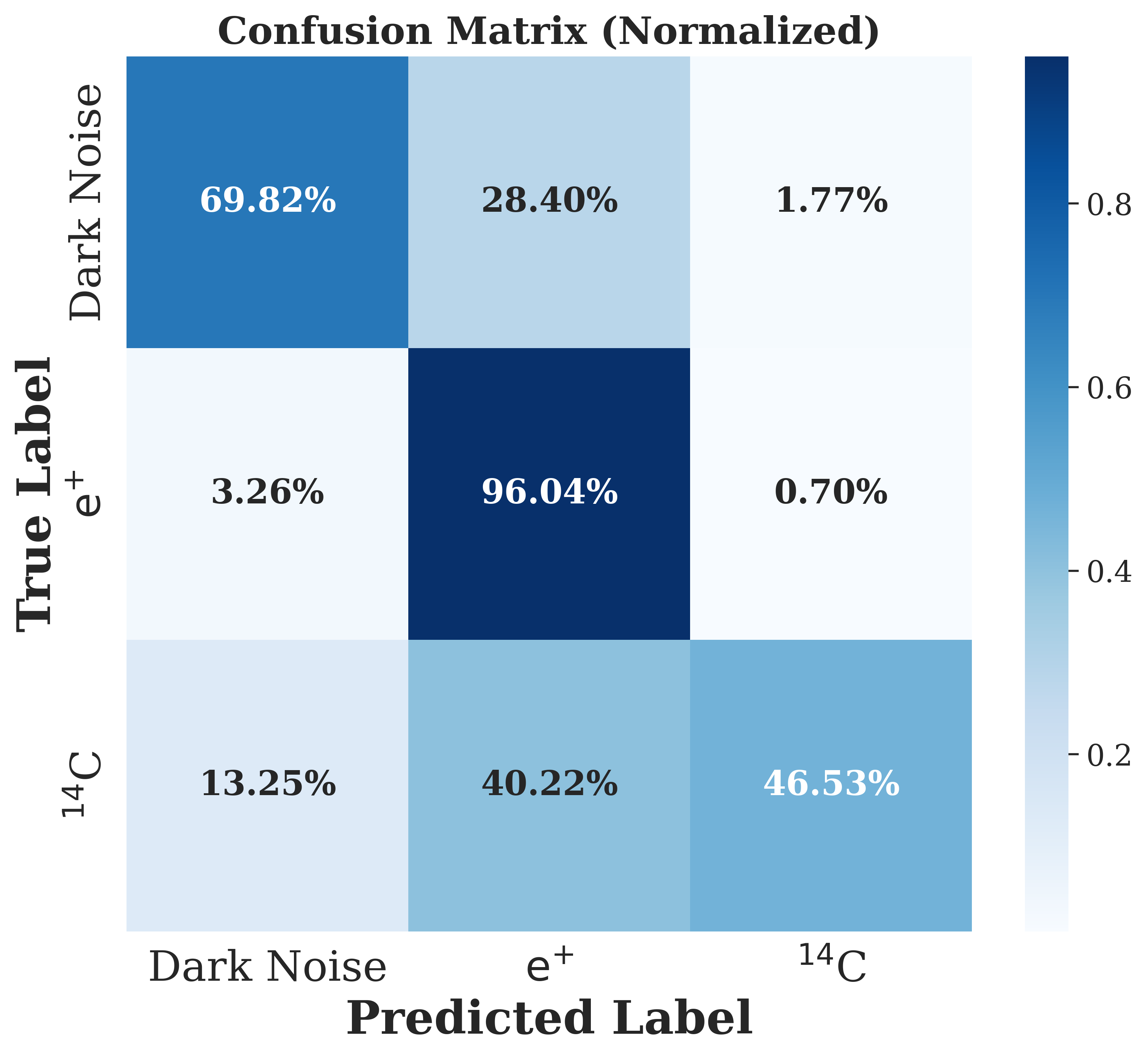}
            
            \vspace{3pt}
            STT-Scalar
            \label{fig:cm_0MeV_scalar}
        \end{minipage}%
    }\hfill 
    \subfloat{%
        \begin{minipage}[b]{0.32\textwidth}
            \centering
            \includegraphics[width=\textwidth]{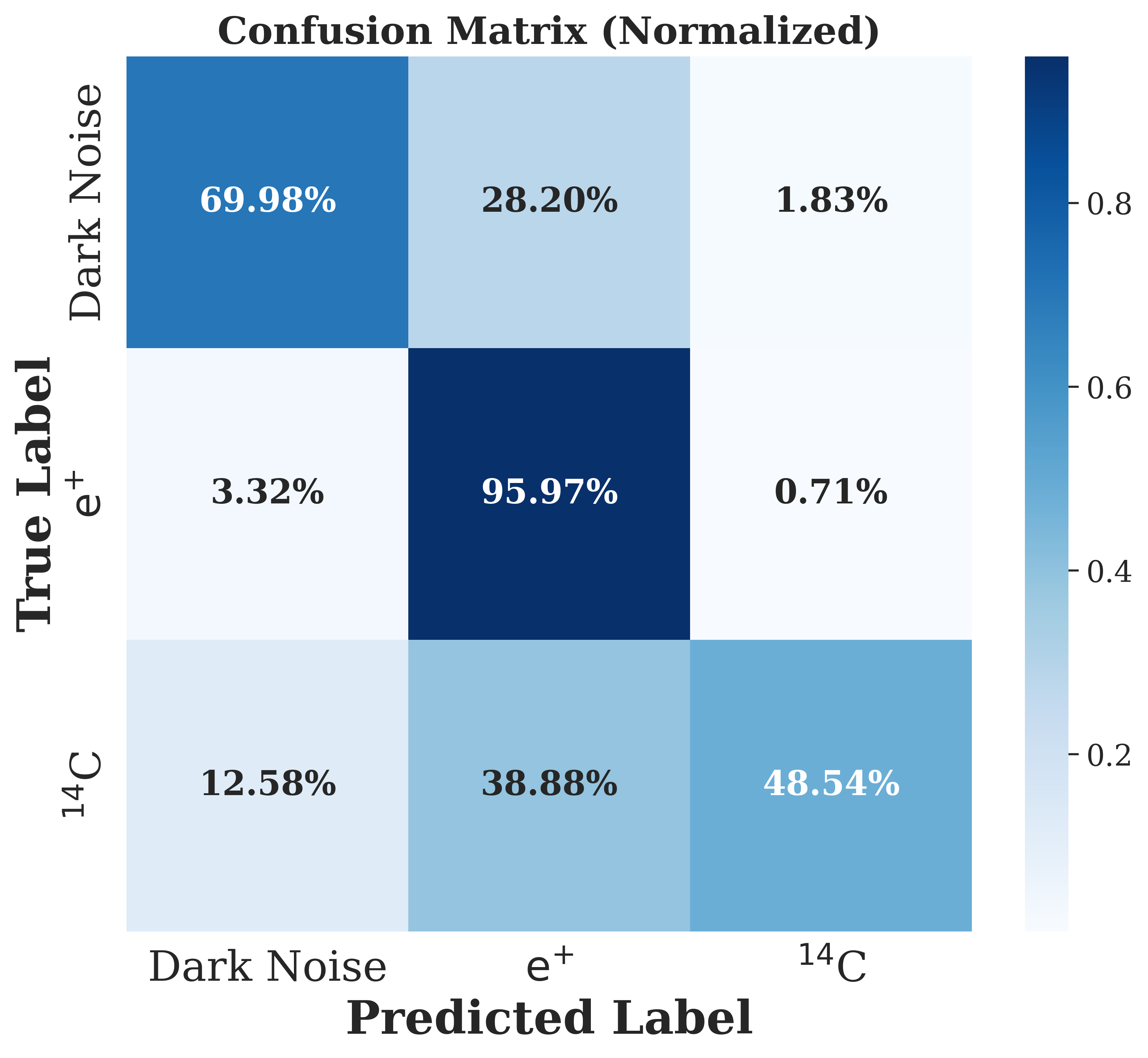}
            
            \vspace{3pt}
            STT-Vector
            \label{fig:cm_0MeV_vector}
        \end{minipage}%
    }

    \caption{Normalized confusion matrices for events with $z=0$ and $E_k({\mathrm{e}^{+}})=0~\mathrm{MeV}$. The panels display the classification performance for  Gated-STGNN, STT-Scalar, and STT-Vector.}
    \label{fig:cm_0MeV}
\end{figure*}

\begin{figure*}[htbp]
    \centering
    
    \subfloat{%
        \begin{minipage}[b]{0.32\textwidth}
            \centering
            \includegraphics[width=\textwidth]{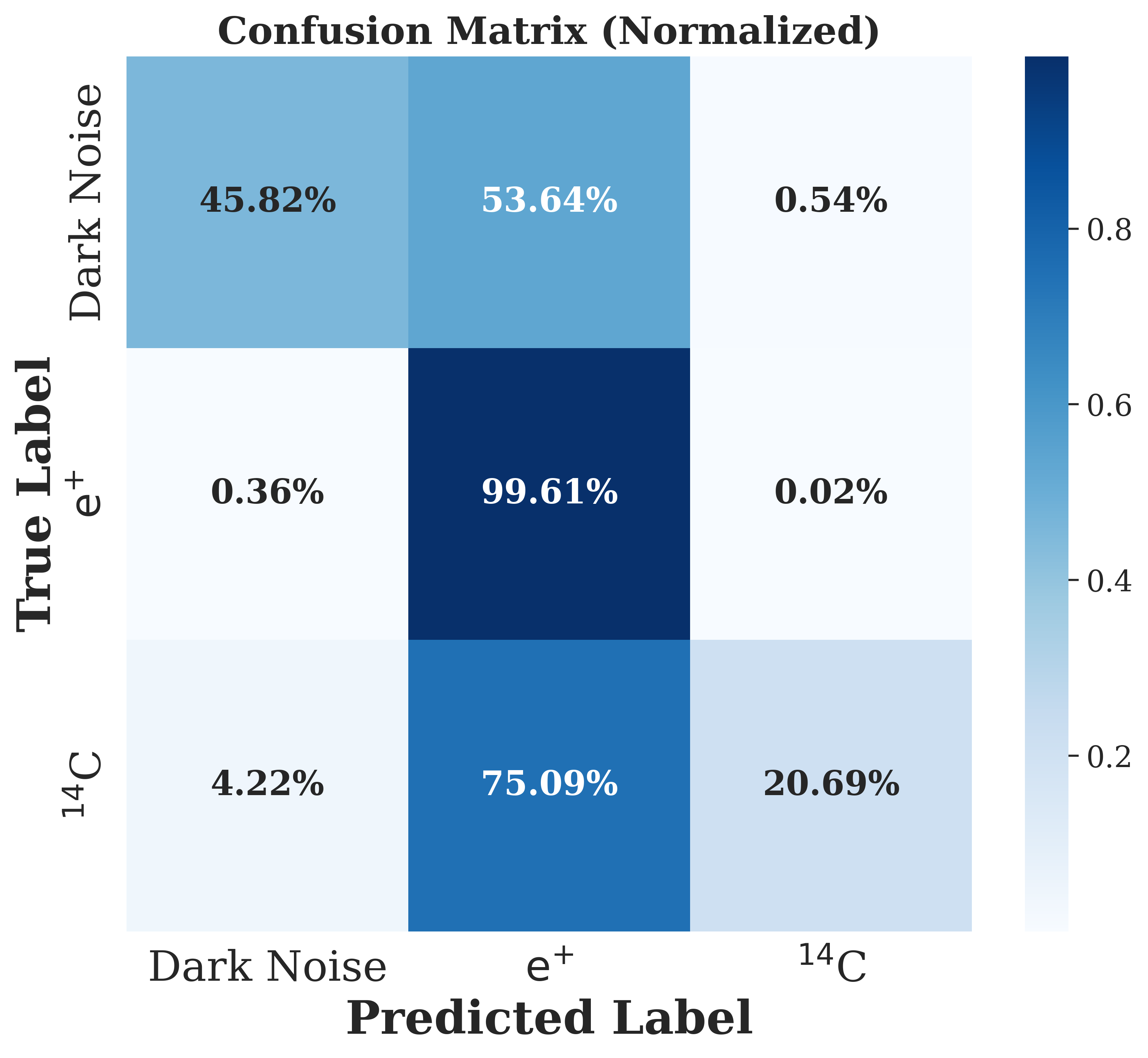}
            
            \vspace{3pt}
            Gated-STGNN
            \label{fig:cm_5MeV_gated}
        \end{minipage}%
    }\hfill 
    \subfloat{%
        \begin{minipage}[b]{0.32\textwidth}
            \centering
            \includegraphics[width=\textwidth]{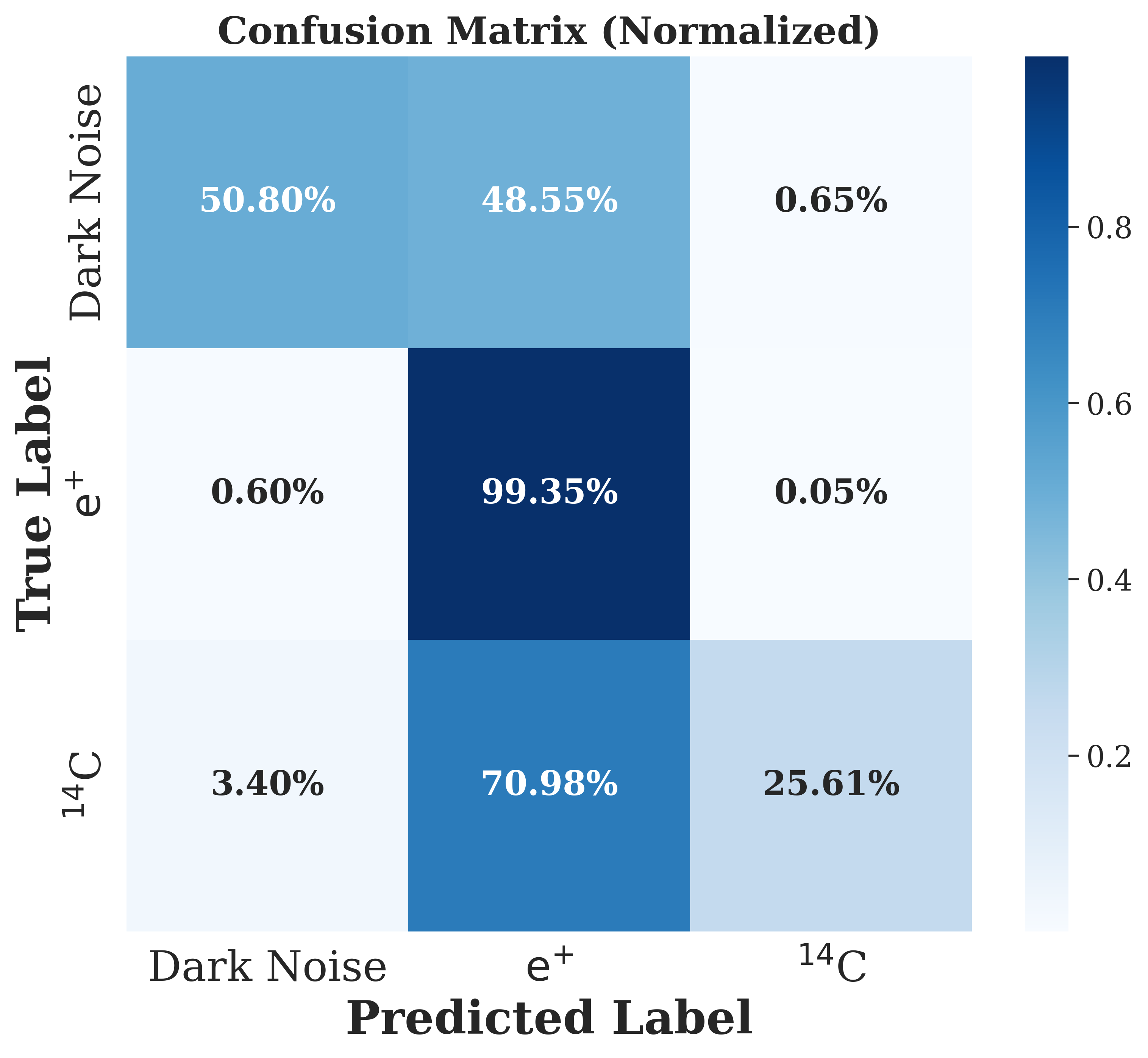}
            
            \vspace{3pt}
            STT-Scalar
            \label{fig:cm_5MeV_scalar}
        \end{minipage}%
    }\hfill 
    \subfloat{%
        \begin{minipage}[b]{0.32\textwidth}
            \centering
            \includegraphics[width=\textwidth]{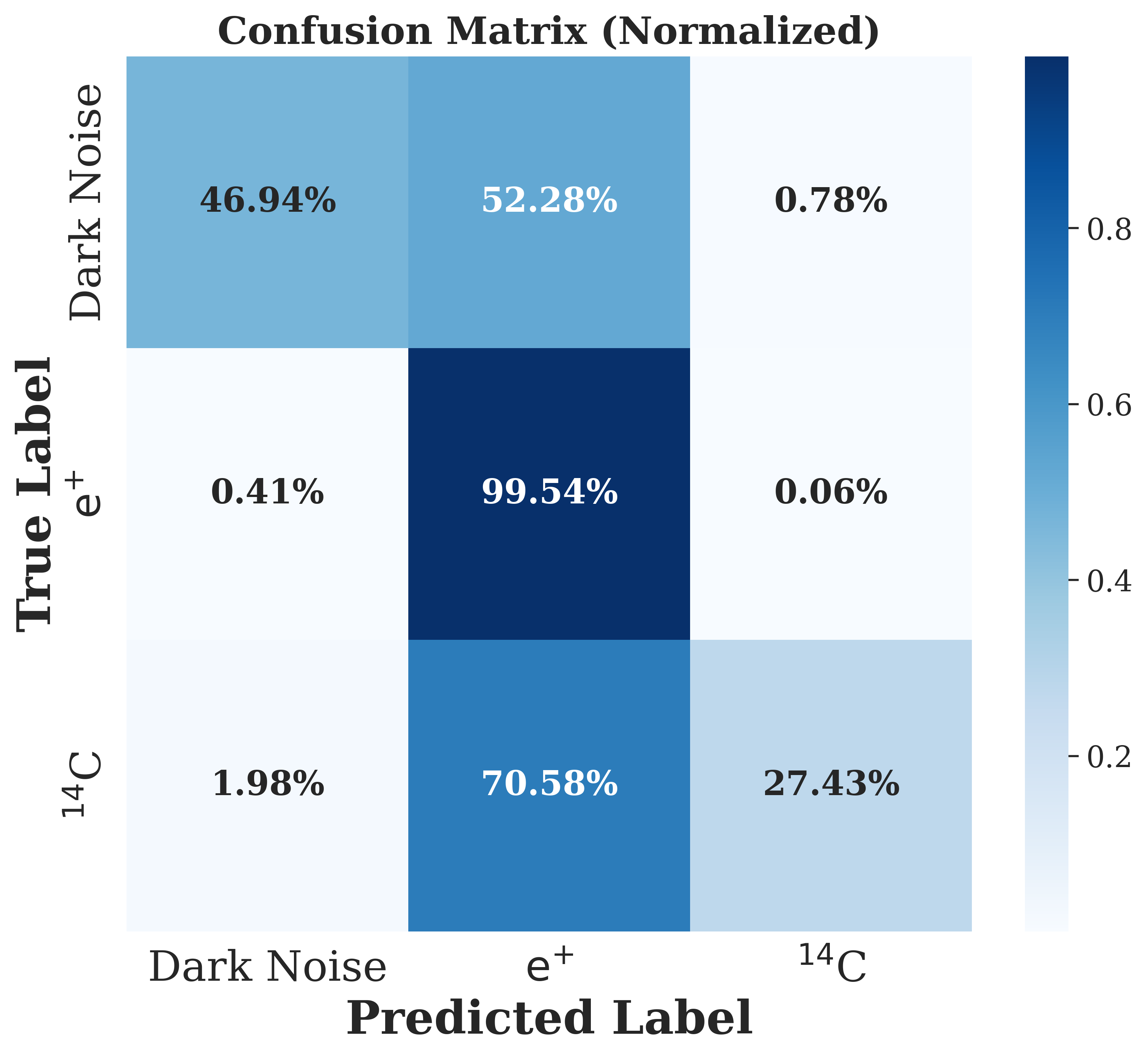}
            
            \vspace{3pt}
            STT-Vector
            \label{fig:cm_5MeV_vector}
        \end{minipage}%
    }

    \caption{Normalized confusion matrices for events with $z=0$ and $E_k({\mathrm{e}^{+}})=5~\mathrm{MeV}$. The panels display the classification performance for Gated-STGNN, STT-Scalar, and STT-Vector.}
    \label{fig:cm_5MeV}
\end{figure*}

\subsection{Training objective}
\label{sec:training_objective}

Three models are trained by minimizing the hit-level cross-entropy loss $\mathcal{L}_{\mathrm{CE}}$ over the set of valid (non-padding) hits $\mathcal{V}$:
\begin{equation}
\mathcal{L}_{\mathrm{CE}} = -\frac{1}{N_{\mathcal{V}}}\sum_{i\in \mathcal{V}}\log p_{i,c_{i}},
\label{eq:ce_loss}
\end{equation}
where $N_{\mathcal{V}}$ denotes the total count of valid hits. The term $p_{i,c_{i}}$ represents the predicted probability assigned to the truth class $c_i$, derived by applying the softmax function to the output logits $\mathbf{s}_i$. The padding masks are applied to exclude invalid entries from the loss computation.

\subsection{Hit-level classification metrics}
\label{sec:hit_metrics}
For each test set, a $3\times 3$ confusion matrix is constructed from the aggregated hit-level predictions. The normalized matrix is defined as
\begin{equation}
C_{ab} = \frac{N(\hat{y}=b,\, y=a)}{\sum_{b' \in \{0,1,2\}} N(\hat{y}=b',\, y=a)},
\end{equation}
where $a$ labels the true class, $b$ labels the predicted class, and $N(\hat{y}=b,\, y=a)$ denotes the number of hits with true label $a$ predicted as $b$. The indices $a, b \in \{0, 1, 2\}$ correspond to the dark noise, positron ($\mathrm{e}^{+}$), and $^{14}\mathrm{C}$ classes, respectively. The diagonal element $C_{22}$ defines the $^{14}\mathrm{C}$ signal efficiency (recall), denoted as $R_{^{14}\mathrm{C}}$. The contamination levels are quantified by the off-diagonal elements $C_{02}$ and $C_{12}$, which represent the misidentification fractions of dark noise ($f_{\mathrm{DN}\rightarrow {^{14}\mathrm{C}}}$) and positron hits ($f_{\mathrm{e}^{+}\rightarrow {^{14}\mathrm{C}}}$) classified as $^{14}\mathrm{C}$. The practical goal is to suppress pile-up from $^{14}\mathrm{C}$ while minimizing the loss of $\mathrm{e}^{+}$ hits, thereby reducing the degradation of the reconstructed energy resolution. To further elucidate model performance, the $^{14}\mathrm{C}$ recall is evaluated differentially as a function of the time separation $\Delta t$, as defined in Eq.~\eqref{eq:delta_t}, and the deposited energy $E_{^{14}\mathrm{C}}$:
\begin{equation}
R_{^{14}\mathrm{C}}(\Delta t, E_{^{14}\mathrm{C}}) =
\frac{N(\hat{y}=2, y=2|\Delta t, E_{^{14}\mathrm{C}})}{N(y=2|\Delta t, E_{^{14}\mathrm{C}})}.
\end{equation}

\begin{figure*}[t]
    \centering
    
    \subfloat{%
        \begin{minipage}[t]{0.32\textwidth}
            \centering
            \includegraphics[width=\linewidth]{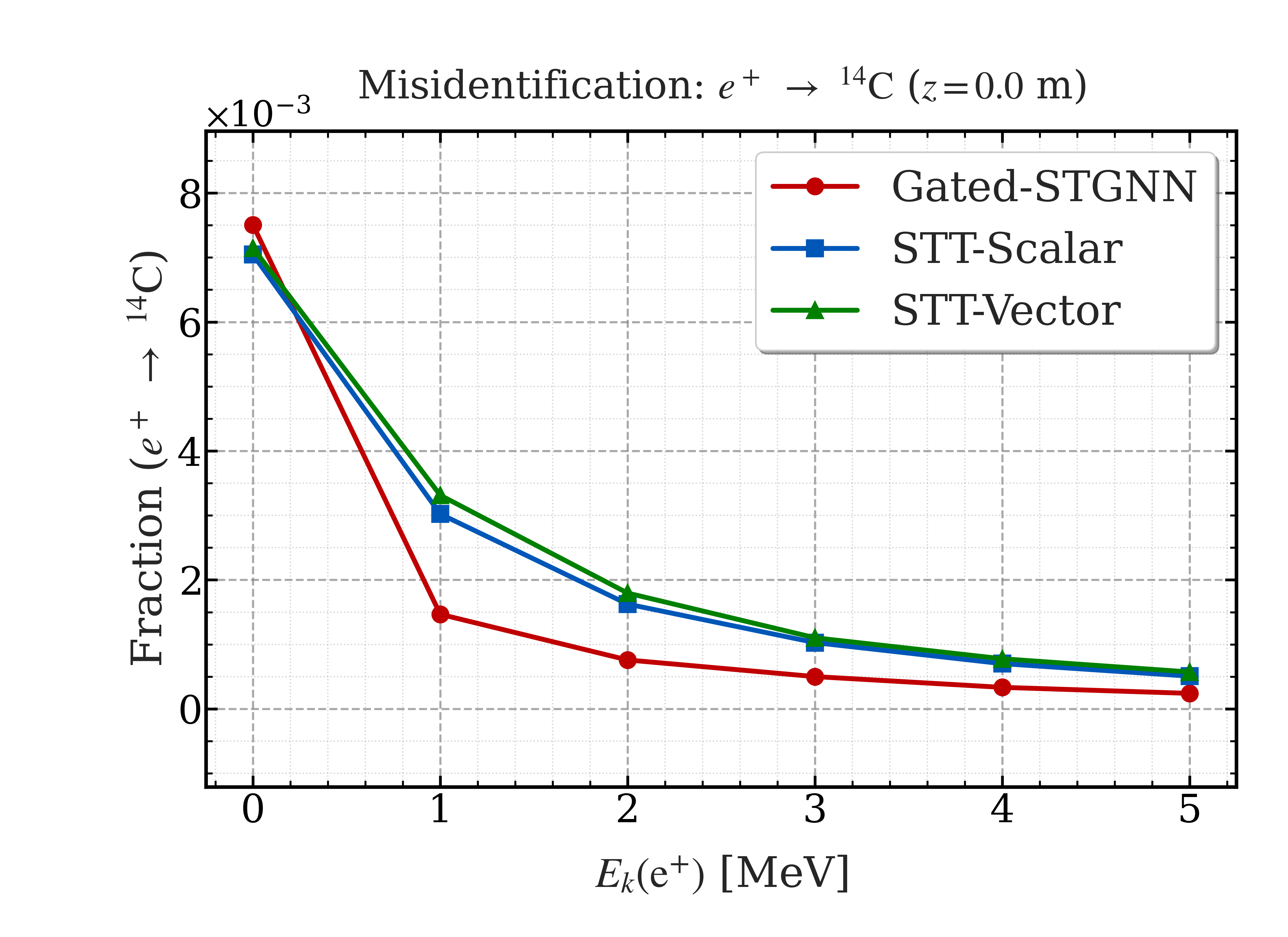}
            
            \vspace{3pt}
            $\mathrm{e}^{+}\rightarrow{}^{14}\mathrm{C}$ Misidentification
            \label{fig:metrics_z0_e_misid}
        \end{minipage}%
    }\hfill 
    \subfloat{%
        \begin{minipage}[t]{0.32\textwidth}
            \centering
            \includegraphics[width=\linewidth]{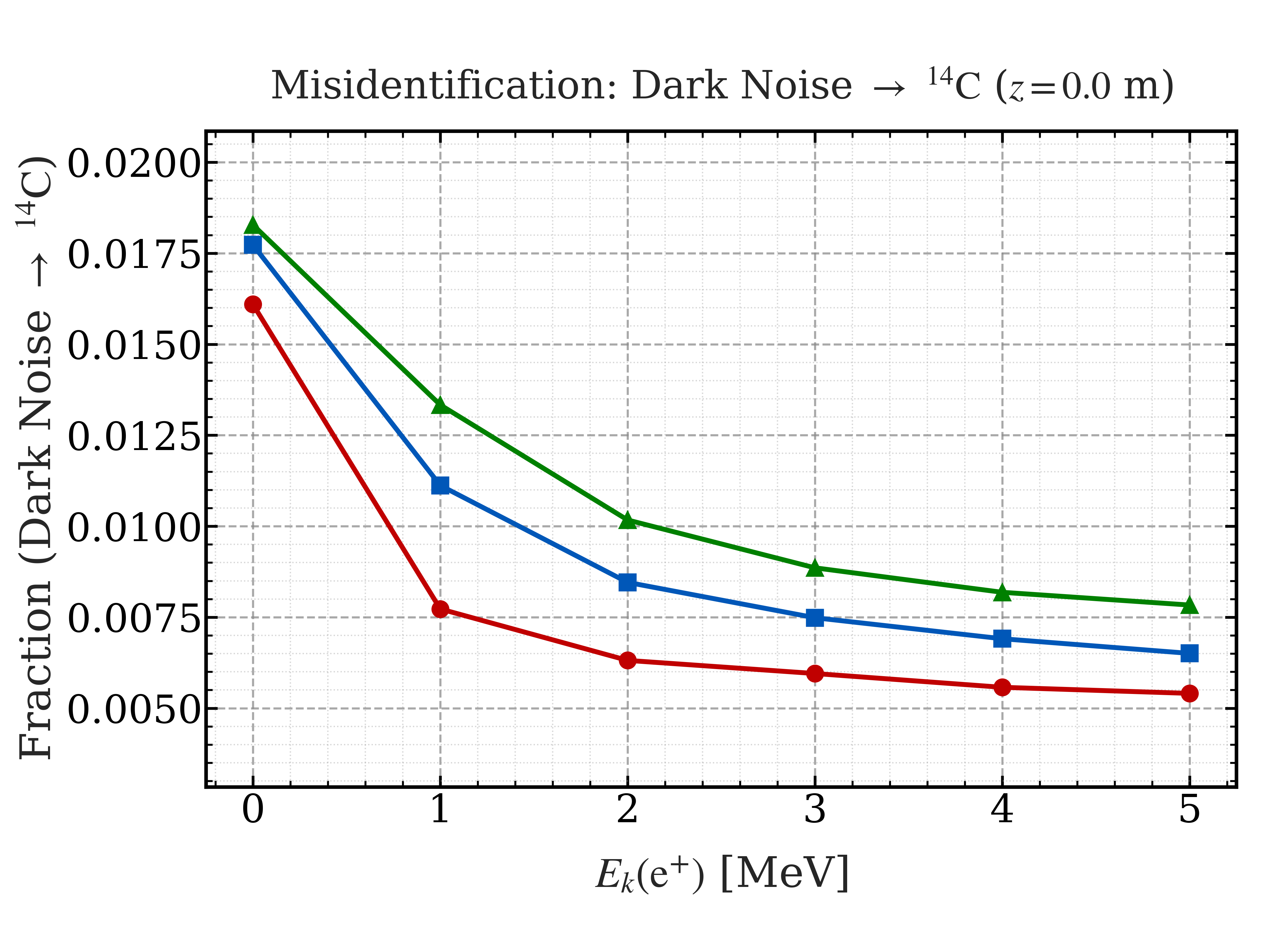}
            
            \vspace{3pt}
            DN$\rightarrow{}^{14}\mathrm{C}$ Misidentification
            \label{fig:metrics_z0_dn_misid}
        \end{minipage}%
    }\hfill 
    \subfloat{%
        \begin{minipage}[t]{0.32\textwidth}
            \centering
            \includegraphics[width=\linewidth]{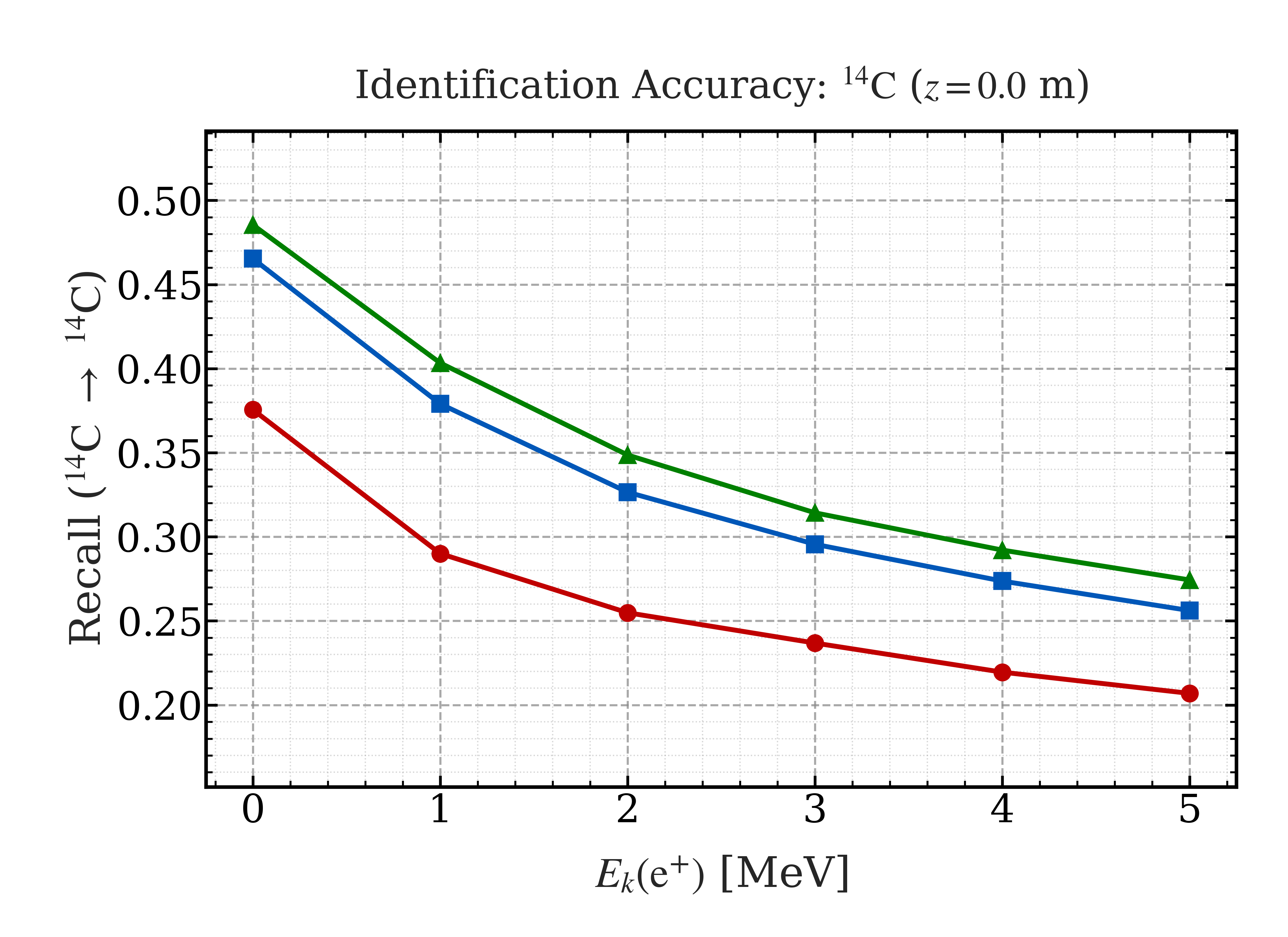}
            
            \vspace{3pt}
            $^{14}\mathrm{C}$ Recall
            \label{fig:metrics_z0_recall}
        \end{minipage}%
    }

    \caption{Classification metrics as functions of positron kinetic energy $E_k({\mathrm{e}^{+}})$ at $z=0~\mathrm{m}$. The panels show the dependence on $E_k({\mathrm{e}^{+}})$ of the fraction of $\mathrm{e}^{+}$ hits misidentified as $^{14}\mathrm{C}$, the fraction of dark noise (DN) hits misidentified as $^{14}\mathrm{C}$, and the $^{14}\mathrm{C}$ recall.}
    \label{fig:metrics_z0}
\end{figure*}

\begin{figure*}[t]
    \centering
    
    \subfloat{%
        \begin{minipage}[t]{0.32\textwidth}
            \centering
            \includegraphics[width=\linewidth]{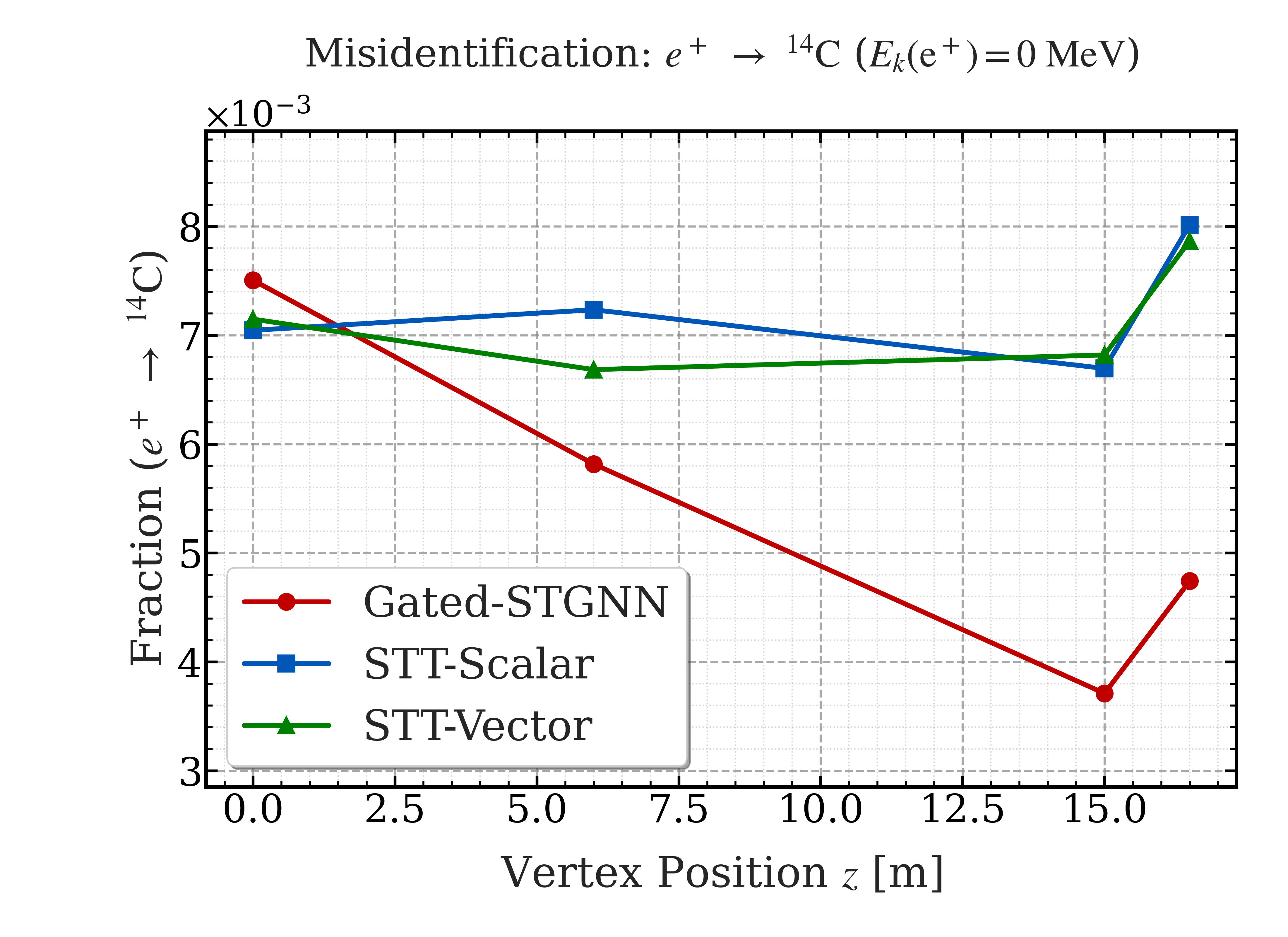}
            
            \vspace{3pt}
            $\mathrm{e}^{+}\rightarrow{}^{14}\mathrm{C}$ Misidentification
            \label{fig:metrics_z_e_misid}
        \end{minipage}%
    }\hfill 
    \subfloat{%
        \begin{minipage}[t]{0.32\textwidth}
            \centering
            \includegraphics[width=\linewidth]{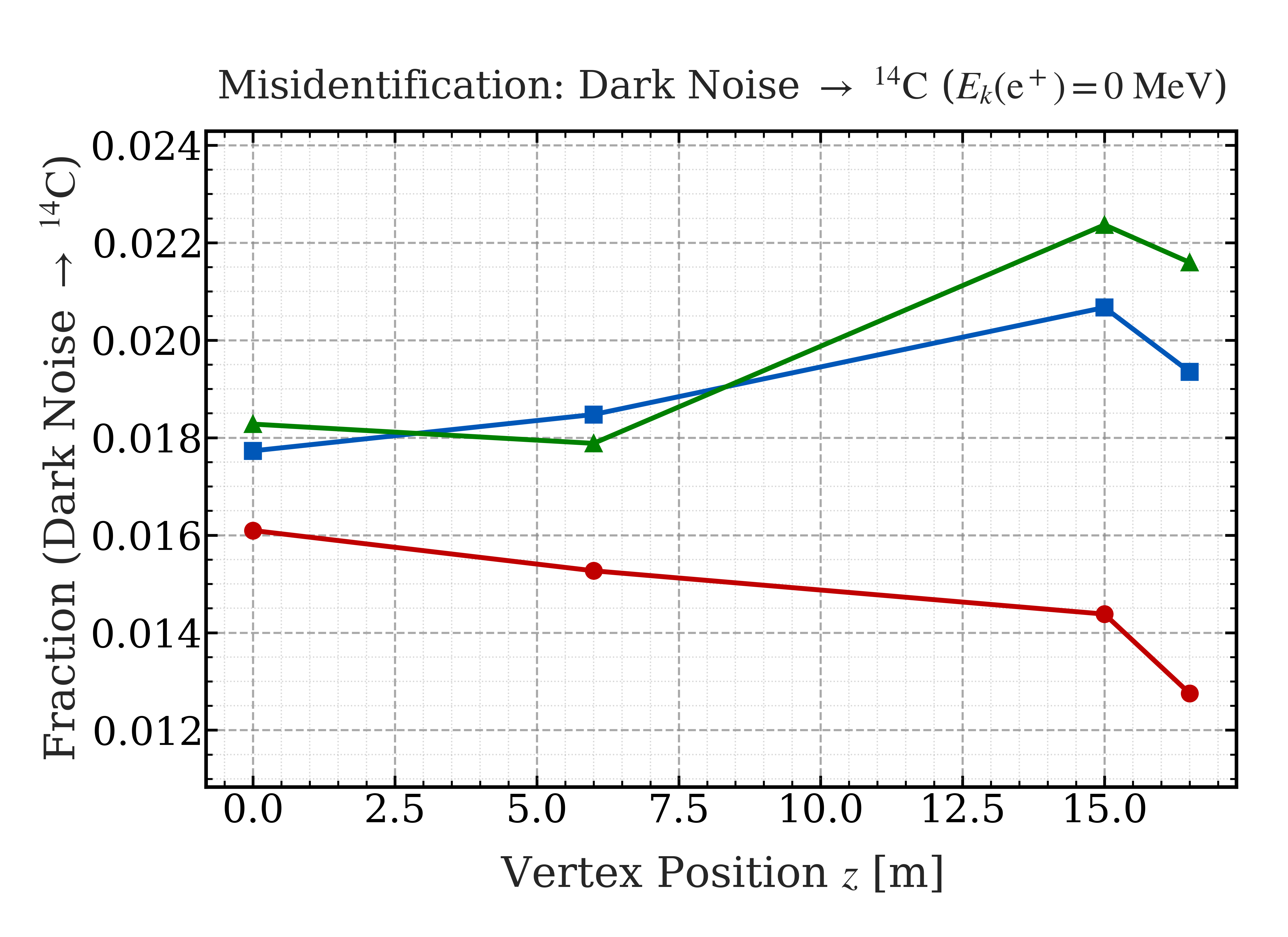}
            
            \vspace{3pt}
            DN$\rightarrow{}^{14}\mathrm{C}$ Misidentification
            \label{fig:metrics_z_dn_misid}
        \end{minipage}%
    }\hfill 
    \subfloat{%
        \begin{minipage}[t]{0.32\textwidth}
            \centering
            \includegraphics[width=\linewidth]{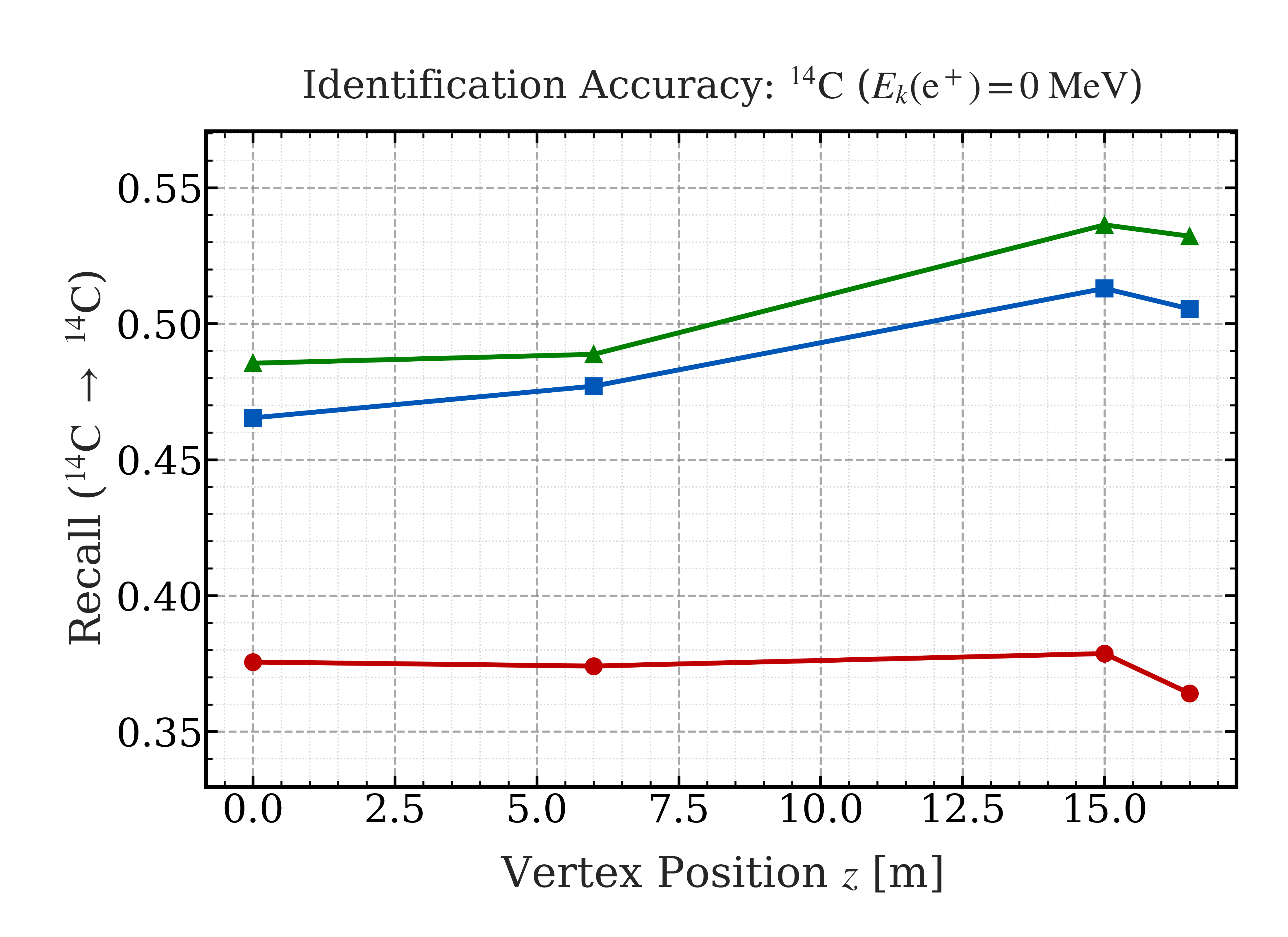}
            
            \vspace{3pt}
            $^{14}\mathrm{C}$ Recall
            \label{fig:metrics_z_recall}
        \end{minipage}%
    }

    \caption{Classification metrics as functions of the vertex position $z$ (evaluated at discrete points $z \in \{0, 6, 15, 16.5\}~\mathrm{m}$) for events with a fixed positron kinetic energy  $E_k({\mathrm{e}^{+}})=0~\mathrm{MeV}$. The panels display the dependence on $z$ of the fraction of $\mathrm{e}^{+}$ hits misidentified as $^{14}\mathrm{C}$, the fraction of dark noise (DN) hits misidentified as $^{14}\mathrm{C}$, and the $^{14}\mathrm{C}$ recall.}
    \label{fig:metrics_vs_position}
\end{figure*}

\begin{figure*}[htbp]
    \centering

    \subfloat[$E_k({\mathrm{e}^{+}})=0~\mathrm{MeV}$\label{fig:heatmap_0MeV_vector}]{%
        \includegraphics[width=0.45\textwidth]{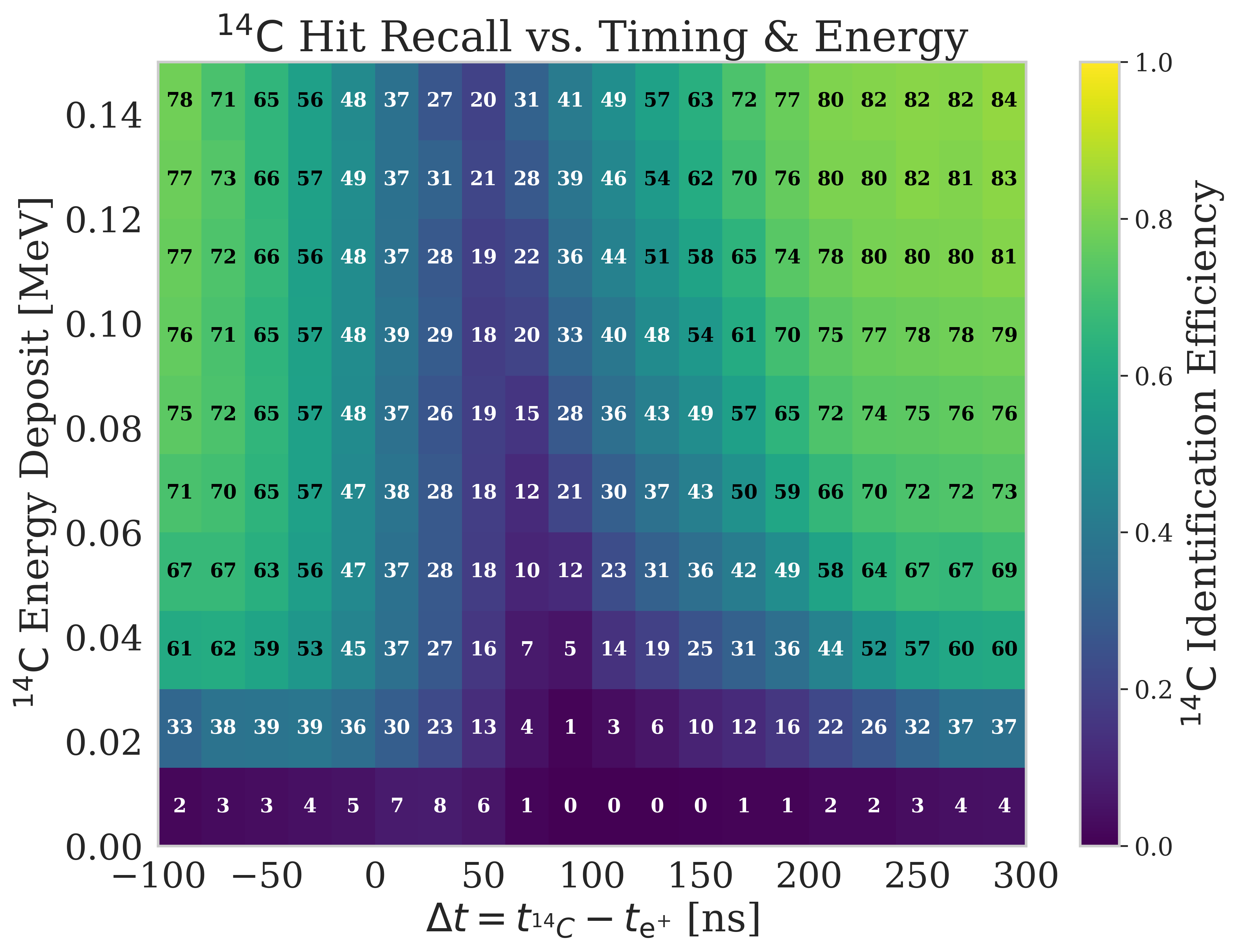}%
    }
    \hfill
    \subfloat[$E_k({\mathrm{e}^{+}})=5~\mathrm{MeV}$\label{fig:heatmap_5MeV_vector}]{%
        \includegraphics[width=0.45\textwidth]{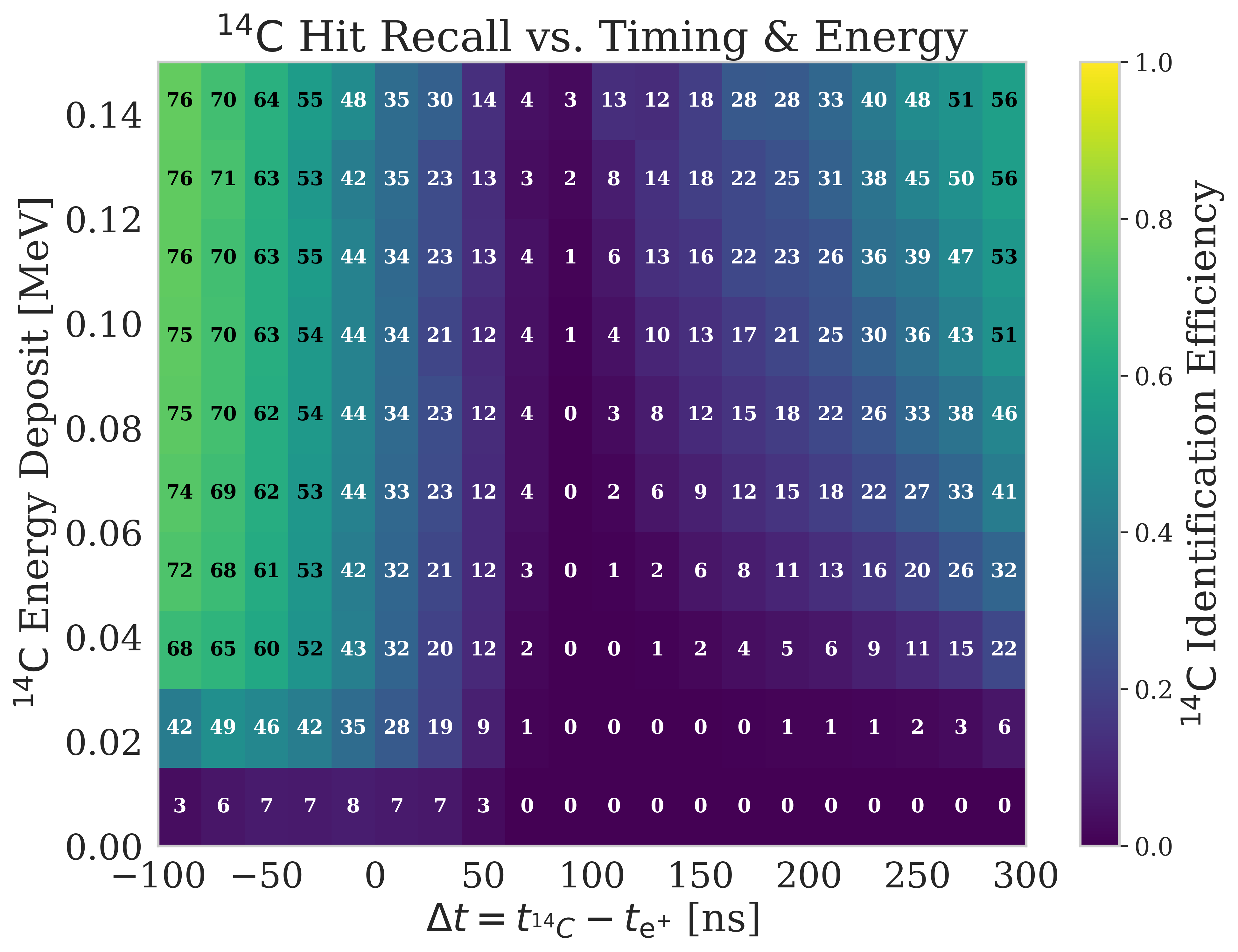}%
    }

    \caption{Differential $^{14}\mathrm{C}$ recall $R_{^{14}\mathrm{C}}(\Delta t, E_{^{14}\mathrm{C}})$ of STT-Vector, evaluated at $z=0$ for $E_k({\mathrm{e}^{+}})=0~\mathrm{MeV}$ and $E_k({\mathrm{e}^{+}})=5~\mathrm{MeV}$, respectively. }
    \label{fig:heatmap_0MeV}
\end{figure*}

\begin{figure*}[t]
    \centering
    \includegraphics[width=0.45\linewidth]{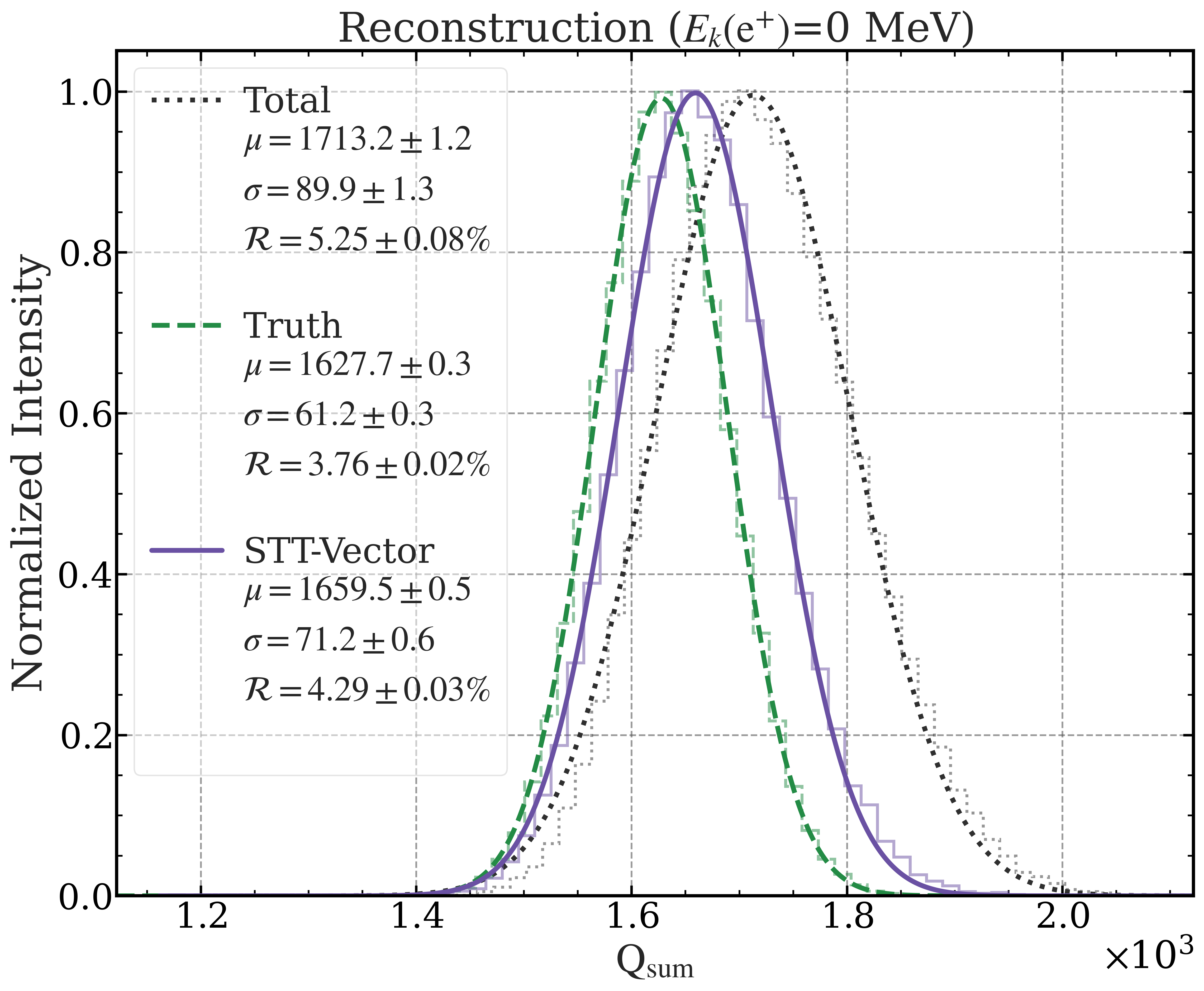}
    \hfill
    \includegraphics[width=0.45\linewidth]{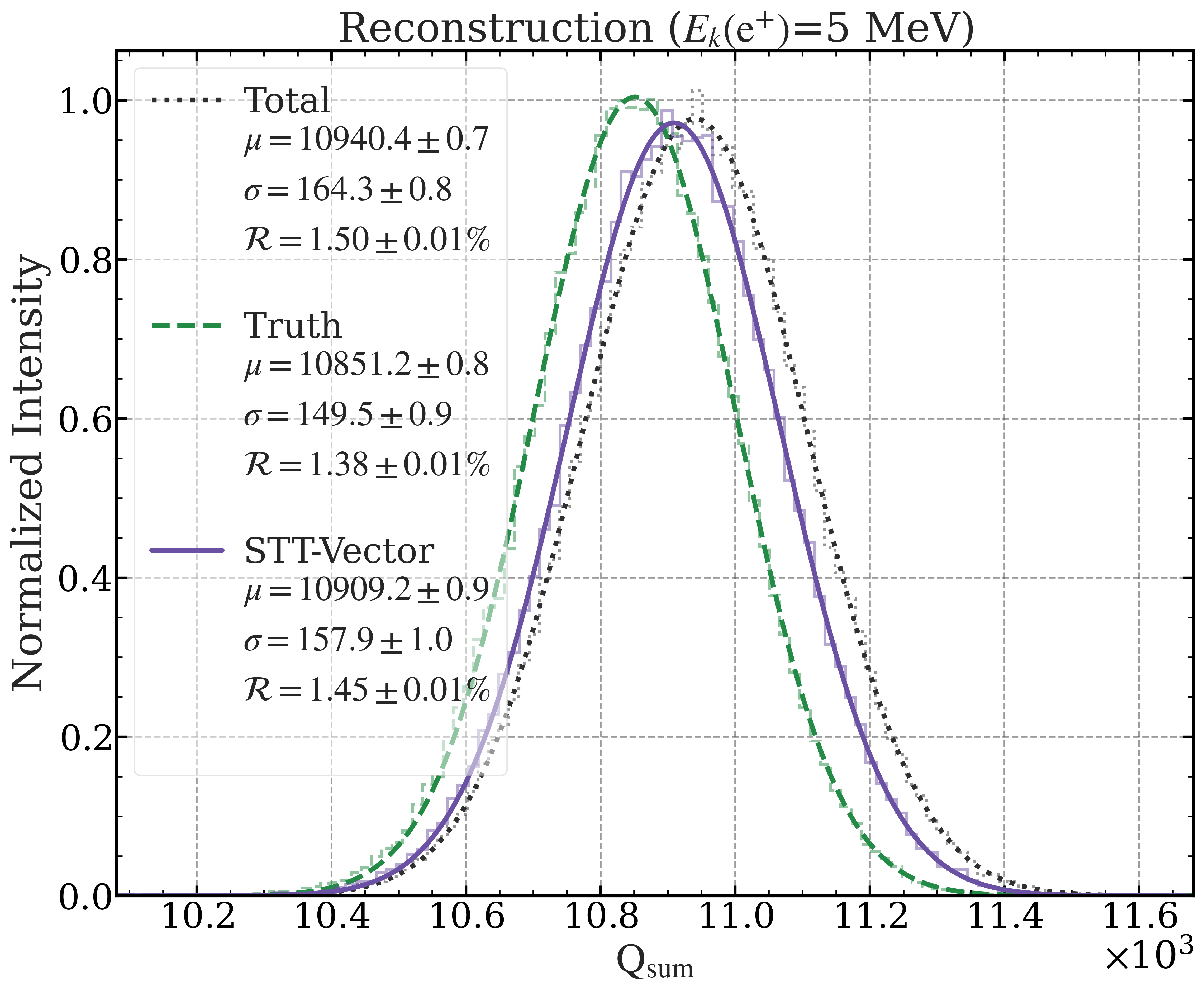} 
    \caption{\label{fig:reco_0MeV} 
    Distributions of the total charge of selected hits ($Q_{\mathrm{sum}}$) at the detector center for events with $E_k({\mathrm{e}^{+}})=0~\mathrm{MeV}$ (left) and $E_k({\mathrm{e}^{+}})=5~\mathrm{MeV}$ (right). 
    The gray dotted line represents $Q_{\mathrm{sum}}$ of all hits, before applying hits removal. 
    The green dashed line shows the ideal distribution with $^{14}\mathrm{C}$ pile-up removed by truth label. 
    The solid purple line represents $Q_{\mathrm{sum}}$ after hit removal based on the STT-Vector model prediction. 
    The fitted mean value ($\mu$), width ($\sigma$) and corresponding resolution ($\mathcal{R} = \sigma/\mu$)  are also shown.
    }    
    \label{fig:dist_0MeV}
\end{figure*}

\subsection{Impact on the Energy Resolution}
\label{sec:resolution_eval}

The removal of $^{14}\mathrm{C}$ hits will reduce fluctuations in the number of detected photons and consequently improve the energy resolution. A simple  study is performed to evaluate the performance of the algorithms in terms of the relative improvement of the energy resolution. This study does not represent the performance of the real JUNO detector, which would require including the improvements from the well-designed energy- and vertex-reconstruction procedure~\cite{JUNO:2024fdc,JUNO:2025fpc}.

The improvement in the energy resolution is quantitatively evaluated using the total charge $Q_{\mathrm{sum}} = \sum_{i} q_{i}$, where $i$ runs over all the selected hits and $q_i$ denotes the charge of the $i$-th hit.

 Hits are selected under different criteria: \par
\textbf{Total} :  all hits are retained. In this case, the $Q_{\mathrm{sum}}$ includes the contributions from $\mathrm{e}^{+}$ and $^{14}\mathrm{C}$, as well as from the dark noise.
The average number of DN hits, $\langle B_{\mathrm{dn}}\rangle=400$, is subtracted from $Q_{\mathrm{sum}}$ to provide a more accurate estimate of the total charge induced by the particles. However, fluctuations in $B_{dn}$ still propagate to the $Q_{\mathrm{sum}}$ resolution.   \par
\textbf{$^{14}\mathrm{C}$ removed (truth)} : $^{14}\mathrm{C}$ hits are removed according to the truth hit labels. In this case, $\langle B_{\mathrm{dn}}\rangle$ is also subtracted from $Q_{\mathrm{sum}}$. It represents the ideal scenario in which the models predict the $^{14}\mathrm{C}$ hits with 100\% efficiency and purity. \par
\textbf{$^{14}\mathrm{C}$ removed (predicted)} : $^{14}\mathrm{C}$ hits are removed according to the hit labels predicted by a model. Since the model will misidentify DN hits as $^{14}\mathrm{C}$ hits, the removal of $^{14}\mathrm{C}$ hits will change the average number of DN hits. Instead, $(1 - C_{02})\langle B_{\mathrm{dn}}\rangle$ is subtracted from $Q_{\mathrm{sum}}$, where $C_{02}$ is the misidentification fraction of dark noise ($f_{\mathrm{DN}\rightarrow {^{14}\mathrm{C}}}$). \par
\textbf{$\mathrm{e}^+$ only (truth)} : only $\mathrm{e}^+$ hits are selected according to the truth hit labels to show the intrinsic resolution of $\mathrm{e}^+$ hits. Since the DN hits are excluded, $\langle B_{\mathrm{dn}}\rangle$ is not subtracted from $Q_{\mathrm{sum}}$.

The $Q_{\mathrm{sum}}$
distributions are fitted with a Gaussian function. The energy resolution is defined as the ratio of the fitted width to the fitted mean, $\mathcal{R} = \sigma/\mu$.

\section{Results and Discussion}
\label{sec:results}

Fig.~\ref{fig:3d_event_display} visualizes the hit-level classification performance for a representative event with $\Delta t = 165.0~\mathrm{ns}$. 
The predictions from the Gated-STGNN, STT-Scalar, and STT-Vector models are consistent with the truth. The models successfully tag $^{14}\mathrm{C}$ hits while effectively discriminating against the dense positron hits and dispersed dark noise hits.

Since light propagation varies significantly with the annihilation vertex,
it is necessary to evaluate the performance of hit-level tagging for $\mathrm{e}^+$ annihilation events at different positions.
At higher positron energies, the relative contribution of $^{14}\mathrm{C}$ pile-up to the total light yield is smaller, 
and its impact on the energy resolution is correspondingly reduced.
However, $^{14}\mathrm{C}$ tagging becomes more challenging at higher positron energies, 
because the $^{14}\mathrm{C}$ hits are statistically overwhelmed by the larger $\mathrm{e}^{+}$-induced hit population, which is typically of order $10^4$. 
Therefore, the identification performance exhibits distinct behavior at low and high positron energies.
The evaluation is conducted across the discrete grid of positron kinetic energies described in Section~\ref{sec:dataset}.
Unless explicitly stated, all models are evaluated on the same set.

\subsection{Global classification performance}

The global classification performance for events with $E_k({\mathrm{e}^{+}})=0$ and $5~\mathrm{MeV}$ at the detector center ($z=0$) is summarized by the normalized confusion matrices in Figs.~\ref{fig:cm_0MeV} and \ref{fig:cm_5MeV}. 
The $^{14}\mathrm{C}$ recall ($C_{22}$) is shown in the bottom-right corner.
Two additional metrics are particularly important: $C_{02}$ and $C_{12}$. They quantify the misidentification fractions of dark-noise hits and positron hits classified as $^{14}\mathrm{C}$, namely $f_{\mathrm{DN}\rightarrow{}^{14}\mathrm{C}}$ and $f_{\mathrm{e}^{+}\rightarrow{}^{14}\mathrm{C}}$. The former biases the effective dark noise level after removal, while the latter reduces the signal light yield. 
Good performance is characterized by high $C_{22}$ and low values of $C_{02}$ and $C_{12}$.

At $E_k({\mathrm{e}^{+}})=0~\mathrm{MeV}$, the STT-Vector model achieves the highest $^{14}\mathrm{C}$ identification recall of 48.54\%, followed closely by STT-Scalar at 46.53\%. In contrast, Gated-STGNN exhibits a notably lower recall of 37.56\%. The primary error mode is the misidentification of $^{14}\mathrm{C}$ as $\mathrm{e}^{+}$, which is 38.88\% for STT-Vector and 47.65\% for Gated-STGNN.
Across all models, the misidentification fraction of $\mathrm{e}^{+}$ as $^{14}\mathrm{C}$ is less than $1\%$ and the misidentification fraction of dark noise as $^{14}\mathrm{C}$ is less than $2\%$. 
This behavior is desirable, indicating that when $^{14}\mathrm{C}$ hits are removed based on the predictions, the signal light yield and the average dark-noise level are barely affected. Consequently, removing the $^{14}\mathrm{C}$ hits will have marginal impact on the downstream event reconstruction.

At  $E_k({\mathrm{e}^{+}})=5~\mathrm{MeV}$, the classification landscape is dominated by the $\mathrm{e}^{+}$ hits. 
Contrary to the low-energy regime, the $^{14}\mathrm{C}$ recall decreases significantly across all models. 
The STT-Vector model maintains the highest recall, achieving a $^{14}\mathrm{C}$ identification recall of $27.43\%$, compared to $25.61\%$ for STT-Scalar and $20.69\%$ for Gated-STGNN. The dominant error mode for all models is the misidentification of $^{14}\mathrm{C}$ hits as positron-induced, with leakage rates exceeding $70\%$. 
This behavior indicates that the $^{14}\mathrm{C}$ hits are statistically overwhelmed by the much larger population of $\mathrm{e}^{+}$ hits,
making their identification increasingly challenging. 
The misidentification rates of $\mathrm{e}^{+}$ as $^{14}\mathrm{C}$ and of dark noise as $^{14}\mathrm{C}$ are both lower than in the low-energy regime. 

The dependence of the $C_{02}$, $C_{12}$ and $C_{22}$ on the positron kinetic energy $E_k({\mathrm{e}^{+}})$ at the detector center ($z=0$) is summarized in Fig.~\ref{fig:metrics_z0}. 
The three metrics show a clear dependence on the positron kinetic energy. As $E_k({\mathrm{e}^{+}})$ increases, both the misidentification rates and the $^{14}\mathrm{C}$ recall decrease rapidly for all three models. For instance, the STT-Vector $^{14}$C-recall curve shown in the right panel (green curve) can be well fitted by an exponential decay function, $A\exp(-E_k/\tau)+C$, with best-fit values of $A=0.3675$, $\tau=2.382~\mathrm{MeV}$, and $C=0.2461$. Figure~\ref{fig:metrics_vs_position} illustrates the  $z$ dependence of the classification metrics with $E_k({\mathrm{e}^{+}})=0~\mathrm{MeV}$. As shown, the STT-Vector model maintains a consistently high $^{14}\mathrm{C}$ recall and low misidentification rates, demonstrating resilience against changes in light path.

\subsection{Timing and energy dependence of \texorpdfstring{$^{14}\mathrm{C}$}{14C} identification}

When the $^{14}\mathrm{C}$ decay occurs close in time to the $\mathrm{e}^{+}$, the hit-level tagging performance  varies strongly with $\Delta t$, since the number of $\mathrm{e}^{+}$ hits surrounding the $^{14}\mathrm{C}$ changes rapidly. To quantify this effect, the $^{14}\mathrm{C}$ hit identification recall is evaluated as a function of the relative time offset $\Delta t$ and the $^{14}\mathrm{C}$ deposited energy $E_{^{14}\mathrm{C}}$.

Figure~\ref{fig:heatmap_0MeV} displays the differential $^{14}\mathrm{C}$ recall as a function of $\Delta t$ and  $E_{^{14}\mathrm{C}}$. 
The maps correspond to events at the detector center ($z=0$) with  positron kinetic energies of $0$ and $5~\mathrm{MeV}$, respectively. For larger $|\Delta t|$, the $^{14}\mathrm{C}$ background hits are temporally decoupled from the prompt $\mathrm{e}^{+}$ signal window, thereby facilitating high $^{14}\mathrm{C}$ recall. In contrast, the region with small $|\Delta t|$ corresponds to strong temporal overlap, where discrimination must rely on subtler correlations among hit timing, geometry, and local charge patterns. When the $^{14}\mathrm{C}$ decays with higher energy, its hits become easier to identify. The overall $^{14}\mathrm{C}$ recall is lower at higher positron kinetic energies, since the larger $\mathrm{e}^{+}$ hit population makes it harder to disentangle $^{14}\mathrm{C}$ hits from the $\mathrm{e}^{+}$ signal.

\begin{figure*}[t]
    \centering
    \includegraphics[width=0.45\textwidth]{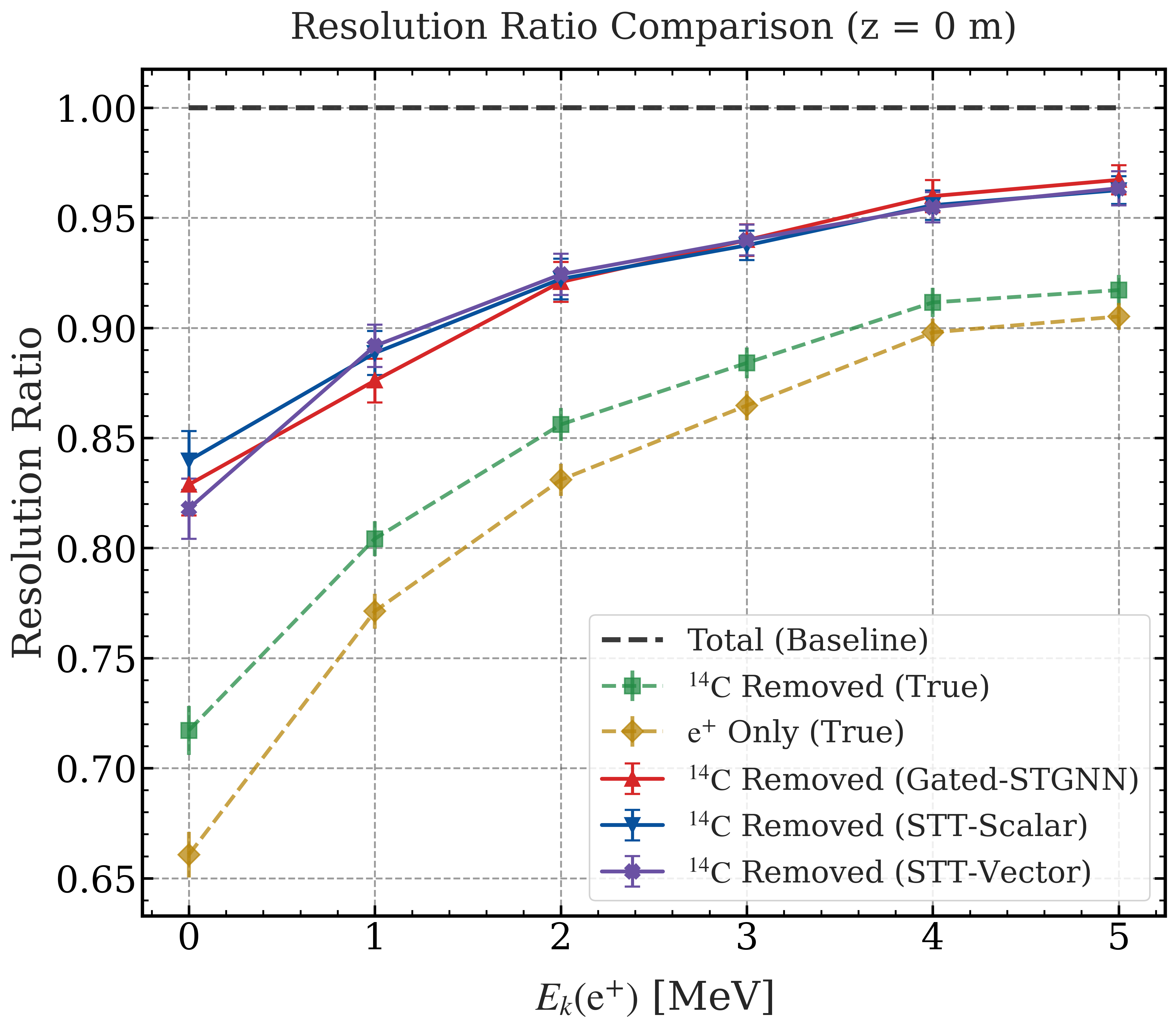}
    \hfill
    \includegraphics[width=0.45\textwidth]{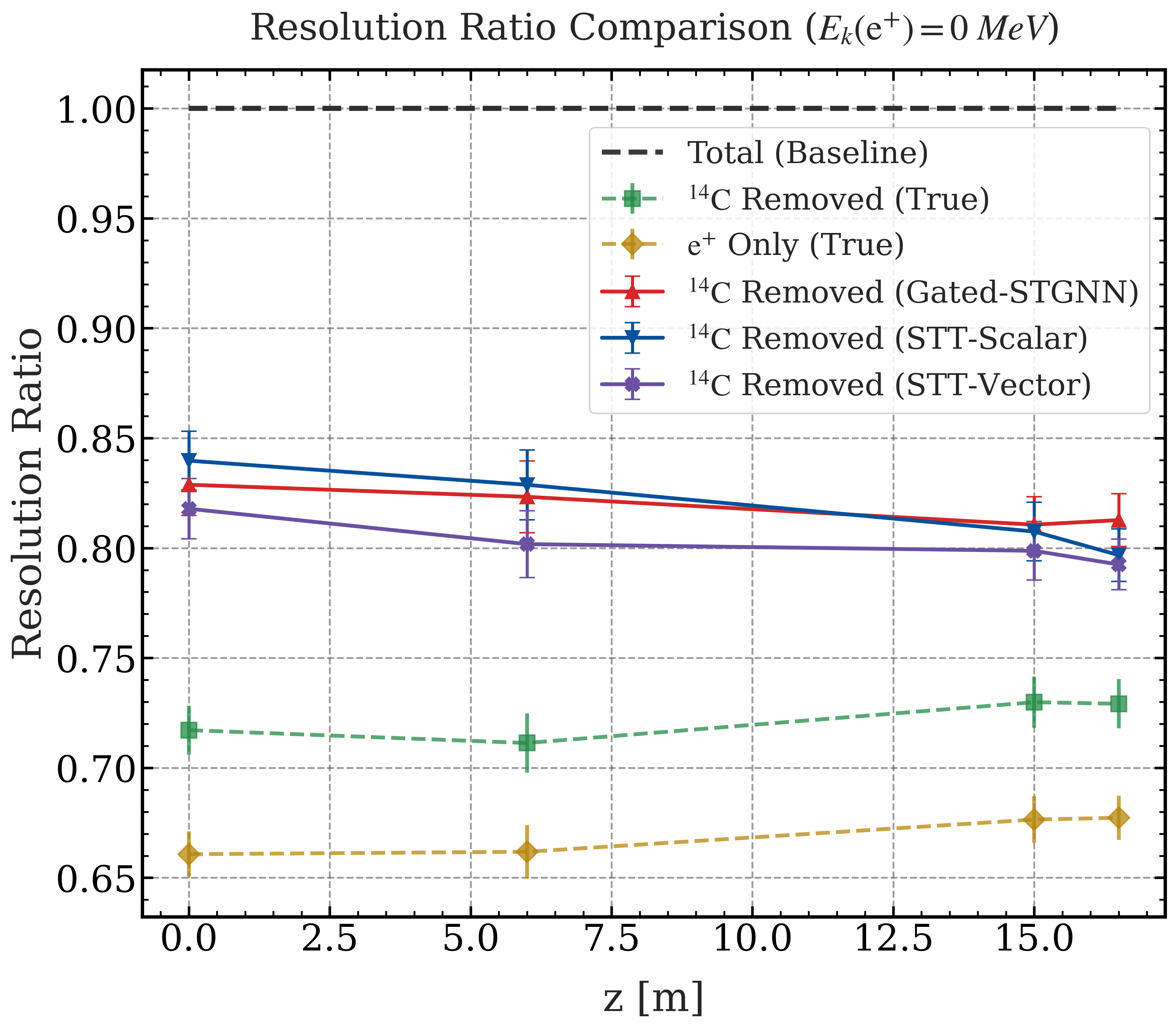}
    \caption{Ratios of the resolutions of the charge sum of selected hits ($Q_{\mathrm{sum}}$) under different hit-removal criteria relative to the ``Total" case, in which no hit is removed. The left plot shows the ratio's dependence on $E_k$ at the detector center. The right plot shows the ratio's dependence on $z$  when $E_k({\mathrm{e}^{+}})=0~\mathrm{MeV}$. }
    \label{fig:res_summary_z0}
\end{figure*}

\subsection{Energy-resolution recovery from hit-level denoising}
\label{subsec:resolution_z0}

The impact of hit-level classification on energy resolution is evaluated through the resolution of $Q_{\mathrm{sum}}$. 
Notably, the strict per-hit classification accuracy does not directly translate to energy resolution performance. For example, consider a positron hit that is spatially and temporally close to a $^{14}\mathrm{C}$ hit.  If a model classifies this hit as a $^{14}\mathrm{C}$ hit, it will be counted as a misidentification, reducing the recall. However, from the perspective of energy reconstruction based on the total charge $Q_{\mathrm{sum}}$ or the likelihood-based reconstruction algorithm~\cite{JUNO:2021vlw}, such a ``mistake" may have negligible impact.   
This is because, in such cases, the overall statistical fluctuation of signal hits has a more significant impact on energy resolution than the precise classification of individual hits.

The $Q_{\mathrm{sum}}$ 
distribution of an $E_k({\mathrm{e}^{+}})=0~\mathrm{MeV}, z=0~\mathrm{m}$ dataset is shown on the left plot of Fig.~\ref{fig:dist_0MeV}. 
After removing $^{14}\mathrm{C}$ hits predicted by STT-Vector, 
the $Q_{\mathrm{sum}}$ resolution improves by $18\%$, approaching the value of the ideal case in which all $^{14}\mathrm{C}$ hits could be removed. In addition, the bias in the mean value of $Q_{\mathrm{sum}}$ with respect to the ideal case is reduced from 5.3\% to 2.0\%. The $Q_{\mathrm{sum}}$ distribution of an $E_k({\mathrm{e}^{+}})=5~\mathrm{MeV}, z=0~\mathrm{m}$ dataset is shown on the right plot of   Fig.~\ref{fig:dist_0MeV}. After removing $^{14}\mathrm{C}$ hits according to the STT-Vector prediction, the $Q_{\mathrm{sum}}$ resolution improves by $3\%$, and the bias in the mean value of $Q_{\mathrm{sum}}$ relative to the ideal case decreases from 0.82\% to 0.53\%.
With higher $E_k({\mathrm{e}^{+}})$, the discrimination of $^{14}\mathrm{C}$ hits becomes more challenging due to the increasing number of $\mathrm{e}^{+}$ hits. 
However, the effect of $^{14}\mathrm{C}$ hits on energy resolution also diminishes, since at high $E_k({\mathrm{e}^{+}})$ the resolution becomes dominated by intrinsic $\mathrm{e}^{+}$ photon statistics. 

The performances of Gated-STGNN, STT-Scalar, and STT-Vector for the datasets with $z=0$ and different $E_k({\mathrm{e}^{+}})$ are summarized on the left of  Fig.~\ref{fig:res_summary_z0}. 
The three models exhibit similar performance. 
In the $E_k({\mathrm{e}^{+}})=0~\mathrm{MeV}$ case, the STT-Vector has better performance. 
As the positron energy increases, the improvement in the $Q_{\mathrm{sum}}$ resolution obtained by removing model-predicted $^{14}\mathrm{C}$ hits diminishes, which is due to both the reduced $^{14}\mathrm{C}$ recall and the smaller fraction of $^{14}\mathrm{C}$ photons at higher energy. In addition, the gap between the model curves and the truth-removal curve (green curve) indicates the remaining room for improvement, particularly in further increasing the $^{14}\mathrm{C}$ recall.
The performances of the three models for the datasets with $E_k({\mathrm{e}^{+}})=0~\mathrm{MeV}$ and different $z$ are summarized on the right of  Fig.~\ref{fig:res_summary_z0}. The behaviors of the three models are similar. An improvement of 16\% to 20\% in the resolution is observed for all tested samples with different $z$, indicating that the proposed models remain valid under more complex light-propagation conditions. 

\section{Conclusion and Outlook}
\label{sec:conclusion}
$^{14}\mathrm{C}$ pile-up contamination is a common and challenging issue for LS detectors such as JUNO. To mitigate its impact on the energy resolution, we have designed a three-stage strategy: machine-learning-based event level $^{14}\mathrm{C}$ pile-up identification, machine-learning-based hit level $^{14}\mathrm{C}$ pile-up identification, and application to the full event reconstruction chain in real data.
In this work, we focus on the second stage and propose three models to tag $^{14}\mathrm{C}$ hits in $\mathrm{e}^{+}$  events affected by a single $^{14}\mathrm{C}$ pile-up. 
By identifying and suppressing background photon hits, the method mitigates the impact of $^{14}\mathrm{C}$ pile-up on event reconstruction. 
The three models are: 
Gated-STGNN, which represents hits as graphs in spacetime and controls information propagation using GRU units; 
STT-Scalar, which applies attention to point-wise features; 
and STT-Vector, which further includes aggregated features
of local topology.

In simulated test datasets with $E_k(\mathrm{e}^{+}) \leq 5~\mathrm{MeV}$, all models successfully tag $^{14}\mathrm{C}$ hits while keeping the misidentification of $\mathrm{e}^{+}$ as $\mathrm{^{14}C}$ below $1\%$ and the misidentification of dark noise as $\mathrm{^{14}C}$ below $2\%$, thereby limiting signal depletion and shifts in the dark-noise level. $^{14}\mathrm{C}$ recall ranges from $25\%$ to $48\%$, 
generally ordered as Gated-STGNN $<$ STT-Scalar $<$ STT-Vector. 
Within the studied overlap regime of $\Delta t \in [-100, 300]\,\mathrm{ns}$, 
differential recall maps indicate that the most challenging region is $\Delta t \in [0, 200]\,\mathrm{ns}$, where the temporal overlap between $^{14}\mathrm{C}$ and $\mathrm{e}^{+}$ is the strongest.

After $^{14}\mathrm{C}$-hit removal based on model predictions, improvements in the resolution of the total charge  $Q_{\mathrm{sum}}$ are observed across all simulated test datasets with different $E_k(\mathrm{e}^{+})$ and $z$ positions, confirming efficacy from the detector center to near-boundary regions under varying positron kinetic  energies.
In the challenging pile-up regime $\Delta t \in [-100, 300]~\mathrm{ns}$, the resolution of $Q_{\mathrm{sum}}$ improves by approximately $5\%$--$20\%$, with STT-Vector yielding the best recovery in the low $E_k({\mathrm{e}^{+}})$ case.

Key limitations include the prioritization of signal retention over aggressive $^{14}\mathrm{C}$-hit tagging, and the dependence on temporal overlap.
Future work should address:
\begin{enumerate}
    \item \emph{Refinement for highly overlapping events:} Incorporating physically motivated descriptors (e.g., time-of-flight corrections, multi-scale context) to improve performance in the $\Delta t \in [0, 200]~\mathrm{ns}$ region.
    \item \emph{Robustness and domain shift:} Systematically validating against various detector conditions and exploring domain-adaptation strategies to mitigate simulation-data discrepancies. In addition to tuning the simulation to better match the data, future work could explore using a mixture of simulated and real data for model training and testing. The residual differences between simulation and data could be treated as systematic uncertainties.
    \item \emph{Integration with reconstruction:} Propagating hit-level selections through the full reconstruction chain to evaluate impacts on the energy and vertex reconstruction.  
    This study focuses on the algorithmic feasibility of hit-level pile-up suppression. 
    The impact on energy resolution is evaluated using $Q_{\mathrm{sum}}$
    as an estimator of calorimetric response at the hit level.
    These results should not be interpreted as a full detector-level performance estimate. 
    A comprehensive assessment requires integration of the hit-selection procedure into the complete JUNO energy reconstruction chain.
\end{enumerate}

\begin{acknowledgments}

We thank Haosen Zhang (Institute of Modern Physics, Chinese Academy of Sciences) for his support with the simulations. This work was partly supported by 
National Natural Science Foundation of China (Grants No. 62472381,
No. 12575209), by the National Key R\&D Program of China (2023YFA1606103).
We would also like to thank the Computing Center of the Institute of High Energy Physics, Chinese Academy of Science and the Earth System Big Data Platform of the School of Earth Sciences, Zhejiang University for providing the GPU resources.

\end{acknowledgments}

% \bibliography{references}

\end{document}